\documentclass[aps,prb,twocolumn]{revtex4-2}

\usepackage{amsmath}
\usepackage{amssymb}
\usepackage{ulem}
\usepackage{color}
\usepackage{graphicx}
\usepackage{epsfig}
\usepackage{subfig}
\usepackage{bm} 
\usepackage[colorlinks=true, linkcolor=blue, citecolor=teal, urlcolor=blue]{hyperref}
\usepackage{natbib}
\usepackage{pgfplots,mathtools}
\usepackage{hyperref}
\usepackage{amsmath}
\usepackage{braket}
\usepackage{slashed}
\usepackage[compat=1.0.0]{tikz-feynman}
\usepackage{physics}
\usepackage{xfrac}
\usepackage{algorithm}
\newcommand{\be}{\begin{equation}}
\newcommand{\ee}{\end{equation}}
\newcommand{\bea}{\begin{eqnarray}}
\newcommand{\eea}{\end{eqnarray}}
\newcommand{\ba}[1]{\begin{array}{#1}}
	\newcommand{\ea}{\end{array}}
\newcommand{\nn}{\nonumber}

\newcommand{\orcid}[1]{\href{https://orcid.org/#1}{\includegraphics[width=8pt]
		{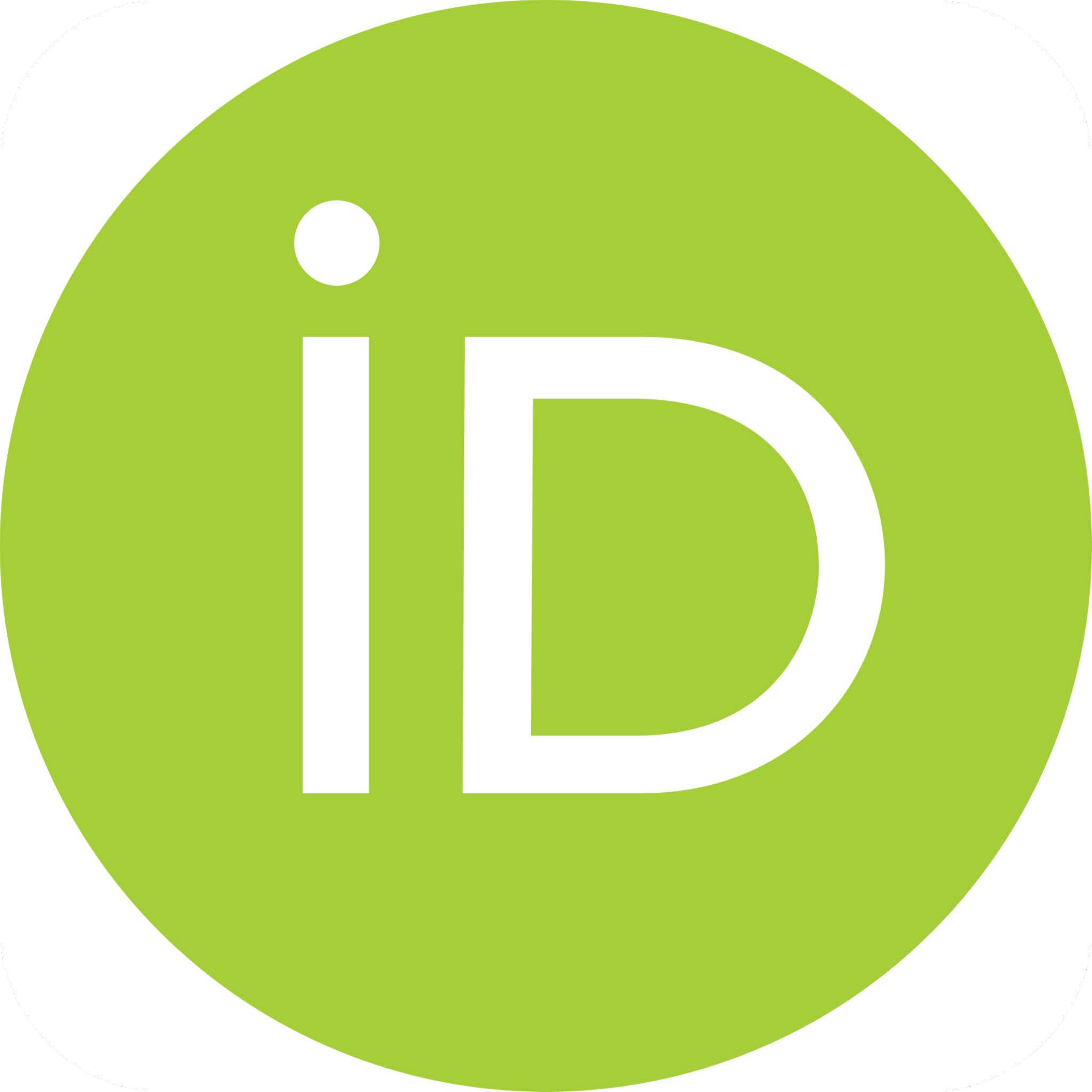}}}
\begin{document}
\title{Probing hydrodynamics in graphene and quark matter via Seebeck coefficient}

\author{Subhalaxmi Nayak\orcid{0009-0001-0145-6785}$^1$}
\email{nayak.subhalaxmi15@gmail.com}

\author{Jayanta Dey\orcid{0000-0002-0894-6402}$^{1,2}$}
\email{jayanta@theor.jinr.ru}
\author{Sabyasachi Ghosh\orcid{0000-0003-1212-824X}$^1$}
\email{sabyaphy@gmail.com}


\affiliation{$^{1}$Department of Physics, Indian Institute of Technology Bhilai, Bhilai, 491002, India.\\
$^{2}$ Bogoliubov Laboratory of Theoretical Physics, Joint Institute for Nuclear Research, Dubna, 141980, Russia.}

\begin{abstract}
We investigate the Seebeck coefficient as a probe of collective transport behavior in graphene and quark-gluon plasma using a kinetic-theory approach in the hydrodynamic regime. The Seebeck coefficient is obtained by solving the Boltzmann transport equation in the relaxation-time approximation.
At high-charge-carrier density, corresponding to the Fermi-liquid domain, our result approaches the behavior expected from the conventional Mott relation. In contrast, significant deviations from the Mott relation are observed in the low-carrier-density regime, corresponding to the Dirac fluid domain. These facts are in good agreement with experimental Seebeck coefficient data for graphene.
This behavior indicates the emergence of collective hydrodynamic transport in graphene. The enthalpy per particle plays a key role for the Seebeck coefficient in the Dirac fluid regime. 
We extend our formalism to the ultra-relativistic quark- gluon plasma. A similar deviation from the Mott relation is observed, supporting a similar fluid response in the Seebeck coefficient across two markedly different strongly correlated quantum systems – graphene and quark matter.

\end{abstract}
                        
\maketitle

\section{Introduction}

Strongly correlated many-body quantum systems are believed to exhibit hydrodynamic behavior when interactions between their microscopic constituents are sufficiently strong. Applicability of hydrodynamics is very vast, from cold atomic gases~\cite{Cao:2010wa} to quark-gluon plasma (QGP)~\cite{Bernhard2019, PhysRevLett.108.252301} and astrophysical objects~\cite{Garbiso2020}. Observable data of quantum chromodynamics (QCD) matter created in heavy-ion collision experiments - Large Hadron Collider and Relativistic Heavy-ion Collider, such as particles' spectra, flow coefficients, etc., are well understood by relativistic hydrodynamics~\cite{DelZaznna2013, PhysRevC.82.014903,jaiswal_AHEP}. As the transport coefficients serve as input parameters for hydrodynamic simulations, many studies, including first-principles simulations of strong-interaction lattice QCD~\cite{PhysRevD.94.034504} and model-dependent studies~\cite{PhysRevD.102.114015, PhysRevC.103.034904,Dwibedi:2024mff,Rai:2025lxw}, and kinetic theory with well-define quasi-particle description~\cite{IJMPE, PhysRevD.109.014018} are used to understand the transport properties of QGP.\\
Hydrodynamic transport is not restricted to relativistic quantum matter. In condensed matter systems, electron hydrodynamics can emerge when momentum-conserving electron-electron scattering dominates over momentum-relaxing processes such as impurity or phonon scattering~\cite{gurzhi1963minimum, Lucasfong2018}.  Experimental signatures of electron hydrodynamics have been reported in graphene~\cite{Bandurin2016,crossno2016observation,sulpizio2019visualizing,Kumar2022}, GaAs~\cite{PhysRevB.51.13389}, PdCoO$_2$~\cite{PdCoO2}, and WP$_2$~\cite{gooth2017electrical}. Graphene is particularly suitable for studying this behavior because its chemical potential can be tuned from eV to a few meV. Its low-energy excitations are massless Dirac quasi-particles with a linear dispersion relation $E(p) = p ~ v_{F}$~\cite{Wallace1947}, where E is the quasi-particle energy, $p$ is the momentum, and  $v_F$ ($\approx 1.1\times 10^6$~m/s) is the Fermi velocity \cite{neto2009electronic}.  The transport regime is governed by the reduced chemical potential $\frac{\mu}{k_{B}T}$, where $k_B$ and $T$ are the Boltzmann constant and temperature, respectively.  For $(\frac{\mu}{k_{B}T}\gg 1)$, graphene is in the Fermi liquid (FL) regime where transport is predominantly diffusive or Ohmic, whereas for $(\frac{\mu}{k_{B}T}\ll 1)$,  strong electron-electron interactions produce collective transport of the carriers in the Dirac fluid (DF) regime \cite{Lucasfong2018,Narozhny2019uib, win2025graphene,Narozhny:2022ncn}. Thus, graphene provides a tunable platform for the investigation of the crossover between conventional and hydrodynamic transport. 
For comprehensive reviews in graphene and its transport properties see Refs.~\cite{Lucasfong2018,neto2009electronic,RevModPhys.83.837,RevModPhys.83.851,nayak2025electronhydrodynamicsgraphene,Narozhny2019uib,Narozhny:2022ncn,Fritz2023Hydrodynamic,Hui_2025}.
In the QGP system, quarks and anti-quarks act as charge and heat carriers. Despite their different microscopic degrees of freedom, strong inter-particle interaction drive both into a regime of collective fluidic transport. A comparative perspective between these high and low-extreme systems is highlighted in Ref.~\cite{Dwibedi2025}.\\
The hydrodynamic regime of graphene exhibits various aspects that deviate from conventional FL expectations, most notably violation of Wiedemann-Franz law in the DF domain~\cite{crossno2016observation,Dwibedi2025,nayak2025electronhydrodynamicsgraphene,win2024wied,majumdar2025universality}. Interestingly, analogous deviations have also been discussed in the QGP domain~\cite{Rath:2019nne,Sahoo:2019xjq,PhysRevD.108.094007,pradhan2023conductivity} as well. These observations suggest that transport coefficients associated with coupled charge and heat currents can serve as a signature of hydrodynamics.\\
Among the thermoelectric transport coefficients, we consider the Seebeck coefficient in the present work, which serves as a useful probe of hydrodynamics.  In a system where charged carriers are responsible for the heat current, a temperature gradient leads to the buildup of an electric potential. The phenomenon is commonly known as the Seebeck effect or thermopower.  The resulting Seebeck coefficient characterizes the coupling between the thermal and electrical transport and is therefore sensitive to the microscopic definition of charge and heat currents. Thermoelectric properties are interesting to study in a system where multiple conserved currents contribute to dissipation processes \cite{ashcroft1993solid}. It therefore provides a possibility to observe the transition from diffusive to collective transport.\\
The thermoelectric response in graphene is strongly enhanced near the Dirac point. Experiments have reported that the Seebeck coefficient for graphene ($S_g$) is of order of ($\approx$$100~ \mu$V/K)  at room temperature, larger than the typical value ($\approx 1\, \mu$V/K) in conventional metals~\cite{PhysRevLett.102.096807, PhysRevLett.116.136802, PhysRevB.108.115418, PhysRevLett.102.166808}. In conventional metals and conductors, the Seebeck coefficient  is commonly described by the semi-classical formulation of Mott~\cite{ashcroft1993solid,PhysRev.181.1336},
\begin{equation}
	S_{\rm Mott}= - \frac{\pi^{2} k_B^{2}T}{3e} \, \frac{1}{\sigma(E)} \frac{\partial \sigma(E)}{\partial E}\Bigg|_{\mu} \label{E10},
\end{equation}
Here $e$ is the elementary charge, $\sigma$ is the electrical conductivity, $\mu$ is the Fermi energy or chemical potential. However, this relation breaks down in clean graphene near the charge neutrality point~\cite{PhysRevLett.116.136802}, where the Seebeck effect rapidly increases with the decreasing gate voltage \cite{PhysRevLett.102.096807,PhysRevLett.102.166808}. It has also been found to be violated in twisted bi-layer graphene~\cite{PhysRevLett.125.226802,Paul2022,ghawri2022breakdown}.\\
Previous studies have reported significant violations of the Seebeck coefficient in graphene from the conventional Mott law, as evidenced by both experiments \cite{PhysRevLett.116.136802,PhysRevLett.102.096807,PhysRevLett.102.166808,doi:10.1073/pnas.1615913113} and theoretical studies \cite{PhysRevB.76.193401,PhysRevB.79.085415,PhysRevB.80.235415,PhysRevB.93.195103,PhysRevB.97.245128,PhysRevB.106.205126, PhysRevB.106.205127}.  Violations have been attributed to quenched disorder, Coulomb interactions, electron-optical phonon scattering, and the slow rate of carrier population imbalance relaxation \cite{PhysRevB.93.195103, PhysRevB.80.235415}. Alternative formulations of the Seebeck coefficient have been developed using energy- dependent scattering time~\cite{PhysRevB.80.235415}, balance equation approaches~\cite{Bao_2010}, and semi-classical methods incorporating band gaps and acoustic phonon scattering~\cite{PhysRevB.86.075411}. A recent study of the Seebeck coefficient has been shown to attain a maximum near the charge neutrality point, depending on the ratio of external and inter-carrier scattering rates, indicating a crossover between the drift diffusive and hydrodynamic transport domains \cite{Atlasov2026}. However, the transition from the conventional Fermi-liquid description to a hydrodynamic expression of the Seebeck coefficient requires further attention to be resolved.\\
In hot and dense quark matter the Seebeck coefficient has been investigated using several theoretical frameworks relevant to heavy-ion collisions phenomenology. These include the Nambu-Jona-Lasinio (NJL), hadron resonance gas (HRG) model, kinetic theory-based relaxation time approximation (RTA) using Boltzmann transport equation (BTE) and Kubo formalism \cite{Abhishek:2020wjm, PhysRevD.99.014015, Das2021,cqsf-537l}. Studies have considered two-flavor and $2+1$ flavor systems as well as the effects of strong and weak magnetic fields~\cite{PhysRevD.102.096011, PhysRevD.104.076021,Singh:2024emy}. These investigations establish the Seebeck coefficients as an important thermoelectric observable in hot and dense QCD matter.\\ 
In this work, we investigate the Seebeck coefficient as a probe of hydrodynamic transport in graphene and massless QGP using kinetic theory framework. We solve the BTE within RTA using a constant relaxation time to obtain the non-equilibrium distribution function arising from electrical and thermal gradients. However, conservation of energy and momentum from RTA in the BTE demands an energy-independent relaxation time, commonly known as the Landau matching condition in hydrodynamics~\cite{landau1959lifshitz}. It is common to consider a thermally averaged relaxation time to overcome this difficulty. Here, we will show that for an energy-independent relaxation time (assuming the same for electron and hole), the Seebeck coefficient for graphene becomes independent of relaxation time. Therefore, the Seebeck coefficient is scale-independent and should exhibit similar behavior across different strongly correlated quantum systems. We study the fluid-based Seebeck coefficient of graphene and compare it with existing experimental observations across different temperatures. We extend our formalism to a single-flavor massless quark-antiquark system and investigate the fluid-based Seebeck coefficient as a function of net quark density. In both cases, we observe a violation of the Seebeck coefficient relative to the conventional Mott relation in the low-density regime, which may indicate the emergence of hydrodynamic behavior.\\
 The structure of this article is arranged as follows. In Sec.~\ref {THEORY}, we derive the Seebeck coefficient for both graphene and quark-gluon plasma within the framework of the kinetic theory by using Boltzmann transport equation under the relaxation time approximation. Sec.~\ref {sec-Results} presents the numerical results, highlighting the dependence of the Seebeck coefficient with respect to the net carrier density. In Sec.~\ref {sum}, we provide a concluding summary of our investigations.
\section{Formalism}
\label{THEORY}
\subsection{Seebeck coefficient for metallic case}
According to classical statistical mechanics by Drude the Seebeck coefficient of electrons in a metal can be expressed as,
 \be
S= -\frac{c_v}{3 n e} = -\frac{k_{B}}{2e} = -0.43 \times 10^{-4} {\rm V/K},  \label{drude-s}
\ee
where $c_v=3nk_B/2$ is the specific heat of electron assumed to behave like an ideal gas. Here $n$ denotes the number density.
However, the observed Seebeck coefficient for metal is of the order of micro-volt per kelvin $(\mu $V/K), which is a factor of 100 times larger than the Drude's estimation~\cite{ashcroft1993solid}. 
This discrepancy was removed by  Sommerfeld using the quantum or Fermi-Dirac statistics for free electron and quantum version of specific heat in the limit of $\frac{\mu}{k_{B}T}>>1$ is obtained as, 
\begin{equation}
	c_v = \frac{n \,\pi^2 \, k_B^2 T}{2 \mu}~. \label{Scv}
\end{equation}
By using Eq.~(\ref{Scv}) in Eq.~(\ref{drude-s}) the Seebeck coefficient is of the following form,
\begin{equation}
	S = -\frac{\pi^2}{6} \frac{k_B^{2}}{e} \frac{ T}{\mu}~,
\end{equation}
which is now a function of $\mu$ and T instead of being a constant.
Later, Mott introduced a more complex structure of Seebeck coefficient by using semi- classical formalism \cite{PhysRev.181.1336,ashcroft1993solid}, which better explains experimental data. The Mott relation is given in Eq.~(\ref{E10}).
The estimated result of the Eq.~(\ref{E10}) can be written as, 
\begin{equation}
	S = -\frac{5\pi^2}{6} \frac{k_B^{2}}{e} \frac{ T}{\mu}. \label{EMOTT_metal}
\end{equation}
Here, we have used degenerate Fermi gas relation $\frac{1}{\sigma(E)} \frac{\partial \sigma(E)}{\partial E}\Bigg|_{ \mu} = \frac{5}{2} \frac{1}{\mu}$.
\subsection{ Hydrodynamic description of Seebeck coefficient in graphene }    \label{gf}
In this section we have given a fluid description for  the  Seebeck coefficient  for the  graphene system within the hydrodynamic regime in the kinetic theory based RTA approach. Description of fluid is understood by two fundamental relation : energy-momentum and charge or number current conservation,
     \begin{equation}
     \partial_\mu T^{\mu \nu} = 0, ~~~~~~~~~~ 
     \partial_\mu j^{\mu} = 0.
     \end{equation} 
Here, $T^{\mu \nu}$ and $j^\mu$ are energy-momentum tensor and current density, respectively. In off-equilibrium system, these quantities can be broken into an equilibrium and some deviation term arises due to dissipation. We express them in terms of fluid degrees for freedom - fluid four-velocity $u^\mu$, energy density $\varepsilon$, pressure $P$ as~\cite{landau1959lifshitz},
\bea
T^{\mu \nu} &=& \varepsilon ~\frac{u^\mu u^\nu}{v_{F}^{2}} - P (g^{\mu \nu} - \frac{u^\mu u^\nu}{v_{F}^2}) + \delta T^{\mu \nu}~, \nn\\
j^\mu &=& n \frac{u^\mu}{v_{F}} + \delta j^\mu .
\eea 
Here $n$ is the net charge carrier density. The off-equilibrium term, $\delta T^{\mu \nu}$ arises due to viscous and thermal dissipation, and $\delta j^\mu$ contributes to electric current dissipation. Here, $g^{\mu \nu}$ is the Minkowski metric tensor.        
    
Now, we describe the system in kinetic theory formalism with a well defined quasi-particle behavior. Transport properties of graphene monolayer can be described considering quasi-relativistic fluid of electrons and holes, exhibiting massless behavior with the dispersion relation $E(p) = p v_F$. The corresponding  equilibrium distribution function for $i^{th}$ fermion species is,
    \be
    f_{i}^{0}= \frac{1}{\exp\beta({E} -  Q_i\mu)+1} ~,\label{E12}
    \ee
    where $Q_i =1~ \text{or}~ -1$ for electrons and holes respectively and $\beta = \frac{1}{k_B T}$.
    Considering free particles, equilibrium thermodynamic quantities can be expressed in the kinetic theory formalism as,
    \bea
    \tilde{n} = g \int \frac{d^{2}p}{h^{2}}~ \left(f_e^0 + f_h^0\right), \nn\\
    n = g \int \frac{d^{2}p}{h^{2}}~ \left(f_e^0 - f_h^0\right), \nn\\
    \varepsilon = g \int \frac{d^{2}p}{h^{2}}~E \left(f_e^0 + f_h^0\right)~, \nn\\
    P = g \int \frac{d^{2}p}{h^{2}}~ \frac{p^2}{2E} \left(f_e^0 + f_h^0\right). 
    \eea
    Here, total and net number density is represented by $\tilde{n}$, and $n$, respectively; $\varepsilon$ is energy density and $P$ is the pressure of the electron-hole system; $g =4$ counts for spin and valley degeneracy. Entropy density ($s$) can be obtain from Euler thermodynamic relation,
    \begin{equation}
    	T s = \varepsilon + P - n\mu~.
    \end{equation} 
Now, to obtain the transport coefficients, we perturbed the system slightly out of equilibrium. The total distribution function in this case can be expressed as,
    \be
    f_{i}= f_{i}^{0} + \delta f_{i} ~,\label{E13}
    \ee
   where $\delta f_i$ is the deviation of distribution function from equilibrium. 
In kinetic theory, the microscopic definition of dissipative electric current density in 2D graphene system is,
      \begin{equation}
      \vec{j} = \sum_{i} q_{i} g \int \frac{d^{2}p}{h^{2}}~ \vec{v}_{F} ~\delta f_{i}  ~.\label{E14} 
     \end{equation}
To obtain $\delta f_{i}$, we solve the BTE under the RTA as~\cite{ANDERSON1974466},
     \begin{equation}
     \frac{\partial f_{i}}{\partial t} + \vec{v}_{F} \cdot \frac{\partial f_{i}}{\partial \vec{x}} + q_{i} \vec{E} \cdot \frac{\partial f_{i}}{\partial \vec{p}} = -\frac{\delta f_{i}}{\tau_i}  ~,\label{E15}
     \end{equation}
     where, $\tau_i$ is the relaxation time of the charge carriers. $\vec{E}$ is the external electric field. Note here that the vector sign over $E$ separate it from the energy which is a scalar quantity.  
     Considering small perturbation, the $\delta f_i$ can be considered up to first order derivative with energy in the Taylor series expansion of distribution function around equilibrium. An ansatz can be taken of the form,
     \begin{equation}
     	\delta f_i = (\vec{v}_F . \vec{\Omega})~ \frac{\partial f_{i}}{\partial {E}}~.\label{E16}
      \end{equation}
     Considering the electric field and temperature gradient responsible for leading the system out of equilibrium, the unknown quantity $\vec{\Omega}$ can be written as,
      \begin{equation}
      	\vec{\Omega}= \alpha_{1} \vec{E} + \alpha_{2} \vec{\nabla} T~.\label{E17}
      \end{equation}
      	 The unknown coefficients $\alpha_{1}$ and $\alpha_{2}$ represent the intensities of the corresponding gradient forces, which drive the system out of equilibrium.
        By solving Eq.~(\ref{E16})  and Eq.~(\ref{E17}) we can obtain the value of $\alpha_{1}$ and $\alpha_{2}$, which can be used to solve the value of $\delta f_i$ (see Appendix ~(\ref{ape1})),
       \begin{equation}
       	 \delta f_{i} =  - q_{i} \tau_i (\vec{v}_{F} \cdot \vec{E})  \frac{\partial f^{0}_{i}}{\partial E} + \frac{{E}- Q_i \mathfrak{h} }{T} \tau_i (\vec{v}_{F} .\vec{\nabla}T ) \frac{\partial f^{0}_{i}}{\partial {E}} ~. \label{E18}
       \end{equation}
       Here $\mathfrak{h}=\frac{\varepsilon+P}{n}$ is the enthalpy per particle, which appears to be an important quantity for Seebeck coefficient of fluid system.\\
       Now Eq.~(\ref{E14}) can be written by using Eq.~(\ref{E18}), as follows,
       \begin{equation}
       	\vec{j}= \sum_{i} \frac{q_{i}g}{2} \int \frac{d^{2}p}{h^2} v^{2}_{F}  \tau_i ~\left[- q_{i} \vec{E} +\left(\frac{{E}- Q_i \mathfrak{h} }{T}\right) \vec{\nabla}T \right]~ \frac{\partial f^{0}_{i}}{\partial E}.  \label{E19}
       \end{equation}
       This current density can carry electric charge $e$ as well as heat. We can define $\vec{j_{e}}$ and $\vec{j_{th}}$ as electric charge and heat current density  respectively. Both can consist of two components - a thermal gradient $\vec{\nabla}T$ and electric field $\vec{E}$. So, they can have proportional relations: 
       Its macroscopic expression can be written as,
       \begin{equation}
       	\vec{j_e}= \sigma_{e} \vec{E} -\kappa_{e}  \vec{\nabla} T= -\sigma_{e} \vec{\nabla} V -\kappa_{e}  \vec{\nabla} T ~, \label{j_e}
       \end{equation}
      \begin{equation}
       	\vec{j_{th}}= \sigma_{th} \vec{E} -\kappa_{th}  \vec{\nabla} T ~.\label{j_th}
       	 \end{equation}
     In matrix form  Eq.~(\ref{j_e}) and Eq.~(\ref{j_th}) can be written as,\\
       	 \begin{equation}
       	 	\begin{pmatrix}
       	 		\vec{j_{e}}  \\
       	 		\vec{j_{th}}
       	 	\end{pmatrix}
       	 	=
       	 	\begin{pmatrix}
       	 		\sigma_{e} & \kappa_{e} \\
       	 		\sigma_{th} & \kappa_{th}
       	 	\end{pmatrix}
       	 	\begin{pmatrix}
       	 		\vec{E} \\
       	 	-\vec{\nabla } T          \label{E20}
       	 	\end{pmatrix}.
       	 \end{equation}
       	 where the proportional constants $\sigma_{e}, \kappa_{e}, \sigma_{th}$ and $\kappa_{th}$, are the corresponding transport coefficients. Among these, $\sigma_{e}$ and $\kappa_{th}$ are called electrical and thermal conductivity respectively.
          The ratio $\frac{\kappa_{th}}{\sigma_{e}T}=L_0$ is known as Lorenz ratio, which remain constant (called Lorenz number $L_{0}=\frac{\pi^2}{3}(\frac{k_B}{e})^{2}=2.44\times10^{-8}~ $W$\Omega/$K$^{2}$ ) for all metals. Indeed, metals always remain in FL or Fermi gas domain $(\frac{\mu}{k_BT}>>1)$ in the thermodynamical phase space, where non-fluid or Ohmic motion of electrons are observed and the Wiedemann-Franz law is followed. However, by reaching towards the DF domain $(\frac{\mu}{k_BT}<<1)$ via tuning the carrier doping in graphene system, Wiedemann-Franz law violation observed experimentally~\cite{PhysRevX.3.041008,crossno2016observation,majumdar2025universality,PhysRevX.3.041008} and theoretically~\cite{win2024wied,Dwibedi2025,PhysRevB.107.085401}. A fluid aspect in DF domain is identified as a possible  reason of Wiedemann-Franz law violation. Let us come to the transport coefficients $\sigma_{e}$ and $\kappa_{e}$, from where Seebeck coefficients can be defined as follows. Reader should notice that the off-diagonal thermal conductivity $\kappa_{e}$ is different from the diagonal thermal conductivity $\kappa_{th}$. The temperature gradient can be built by the electric field even when there is no electric current i.e., $\vec{j_e}=0$,  
       \begin{align}
       0 &= \sigma_{e} \vec{E} - \kappa_{e} \vec{\nabla}T, \nn\\
       \vec{E} &= \frac{\kappa_{e}}{\sigma_{e}} ~\vec{\nabla} T = S_g \vec{\nabla} T, \label{E21}
       \end{align}
where the Seebeck coefficient for graphene is defined as
      \begin{equation}
      S_g=\frac{\kappa_{e}}{\sigma_{e}} ~,\label{E21a}
      \end{equation}
with
      \bea
      	\sigma_e &=&-\sum_{i} \frac{q^{2}_{i}g}{2} \int \frac{d^{2}p}{h^2} v^{2}_{F}  \tau_i\frac{\partial f^{0}_{i}}{\partial {E}} \nn\\
      	&=& \sum_{i} \frac{q^{2}_{i}g}{2}\beta \int \frac{d^{2}p}{h^2} v^{2}_{F}  \tau_i f^0_i (1-f^0_i),
       \label{E22}
       \eea
 and 
       \bea
	\kappa_{e} &=& -\sum_{i} \frac{q_{i}g}{2 T} \int \frac{d^{2}p}{h^2} v^{2}_{F}  \tau_i ~\left(E- Q_i\mathfrak{h}\right) ~ \frac{\partial f^{0}_{i}}{\partial E} \nn\\
	&=& \sum_{i} \frac{q_{i}g}{2} \beta^2 \int \frac{d^{2}p}{h^2} v^{2}_{F}  \tau_i ~\left(E- Q_i\mathfrak{h}\right) ~ f^0_i (1-f^0_i). ~~\label{E23}
	\eea
 Index $`i'$ in summation stands for electron and hole. Considering the same relaxation time for electron and hole, the expression of Seebeck coefficient can be written from Eq.~(\ref{E21a}), (\ref{E22}), and (\ref{E23}) as
\begin{equation}
	S_g = \frac{\sum_{i} q_{i} \int d^{2}p ~\left(E- Q_i\mathfrak{h}\right) ~ f^0_i (1-f^0_i)}{T\sum_{i} q^{2}_{i} \int d^{2}p \, f^0_i (1-f^0_i)}.
\end{equation}	
Now, towards the charge neutrality point ($\mu \rightarrow 0$), $\sigma$ can be analytically solved as follows. Substituting charge of electron and hole, $q_i = \mp e$, and the same constant relaxation time for both, as $\tau_i = \tau_R$, we get
\bea
\sigma_e &=& \sum_{i} \frac{q^{2}_{i}g}{2} \beta \int \frac{d^{2}p}{h^2} v^{2}_{F}  \tau_R f^0_i (1-f^0_i) \nn\\
&=& \frac{g\, 2\pi \beta e^2 \tau_R}{h^2} \int_{0}^{\infty} \frac{E \, e^{\beta E}}{\left(e^{\beta E}+1\right)^2} dE \nn\\
&=& \frac{g\, 2\pi \beta e^2 \tau_R}{h^2} \frac{\partial}{\partial \beta} \left(\int_{0}^{\infty} \frac{1}{\left(e^{\beta E}+1\right)} dE \right)\nn\\
&=& \frac{g\, 2\pi \beta e^2 \tau_R}{h^2} \frac{\eta(1) \Gamma(1)}{\beta^2}\nn\\
&=& \frac{4e^2}{h} \phi_\sigma,
\label{sigma0}
\eea
where $\phi_\sigma = 2\pi \frac{\tau_R}{h\beta} \ln{2}$, is a dimensionless quantity, and it is proportional to temperature. However, $\kappa_{e}$ diverges at $\mu = 0$, because enthalpy per particle, $\mathfrak{h}$ diverges. Therefore, in this formalism, Seebeck coefficient diverges towards the charge neutrality point.\\   
Also, one can simplify the expressions for electrical conductivity, thermoelectric response coefficient and Seebeck coefficients in terms of Fermi integral function as (see Appendix~(\ref{ape2}))
       \begin{equation}
      	\sigma_e= \frac{ e^{2}g \pi \tau_{R} k_{B} T}{h^2} ~\{f_{1}(A)+f_{1}(A^{-1}) \}~.\label{E27}
      \end{equation}
    Here $f_{1}(A) ~\text{and} ~f_{1}(A^{-1})$ are the Fermi integral functions for electrons and holes.
    
    The general form of Fermi integral function for electrons and holes have the following form :
    \bea
    f_{\nu}(A) &=&\frac{1}{\Gamma(\nu)}\int\frac{x^{\nu-1}}{ A^{-1}e^x+1} dx, 
    \nn\\
    f_{\nu}(-A) &=&\frac{1}{\Gamma(\nu)}\int\frac{x^{\nu-1}}{ Ae^x+1} dx, \label{E28}
    \eea
    with $\nu=0,1,2,...$ and $A=exp(\frac{\mu}{k_B T})$ and $\Gamma(\nu)= \int_{0}^{\infty} x^{\nu-1} e^{-x} dx$.
Similarly, solving the Eq.~(\ref{E23}) we get (see appendix~(\ref{ape2})),
      \bea
      	\kappa_{e} &=& - \frac{e~ g ~\pi~ \tau_{R} ~k^{2}_{B}~T}{h^2}  \Big[ 2 \{f_{2}(A)-f_{2}(A^{-1})\} \nn\\
      	&& ~~~ -\mathfrak{h} \beta\{f_{1}(A)+f_{1}(A^{-1})\}  \Big] ~.\label{E29}
      \eea
      Using Eq.~(\ref{E27}) and (\ref{E29}) in Eq.~(\ref{E21a}) we get the Seebeck coefficient in the graphene system in terms of the Fermi integral function as,
     \begin{equation}
     S_g =\frac{k_B}{e}\left[  2 \frac{f_{2}(A)-f_{2}(A^{-1})}{f_{1}(A)+f_{1}(A^{-1})} - \mathfrak{h} \beta\right]~.\label{E32}
     \end{equation}
    
     \subsection{Seebeck coefficient for quark-gluon-plasma} \label{QGP}
    An interesting parallel between electron hydrodynamics in graphene and quark hydrodynamics in quark matter has been highlighted by our earlier works \cite{Dwibedi2025,win2025graphene,Aung:2023vrr,Aung:2025cbo}.
     In particular Ref.~\cite{Dwibedi2025} demonstrated the transport coefficients within the framework of relativistic hydrodynamics where we also showed the violation of the fundamental laws in the QGP system due to the hydrodynamic behaviour of the constituent quarks. Here in this article we have shown how Seebeck coefficient varies with the chemical potential and number density. Several works for Seebeck coefficients has been done for QGP system using different theoretical models \cite{Singh:2024emy,Das2021} and the references are there in.
     We have derived an expression for Seebeck coefficient by using RTA based formalism. To reduce complexity we have taken a single flavored massless quark and antiquark ($q$, $\bar{q}$) system with total degeneracy $g=6$  $i.e.,$ color degeneracy $3 \times$ spin degeneracy $2$ for reference. As mentioned in our earlier work Ref.~\cite{Dwibedi2025} we have used ($q$, $\bar{q}$) as ($u$, $\bar{u}$). The local distribution functions for ($u$, $\bar{u}$) are $f^{0}_{u}=1/(e^{(E_{u}-\mu_{u})/k_{B}T}+1)$ and $f^{0}_{\bar{u}}=1/(e^{E_{\bar{u}}-\mu_{\bar{u}})/k_{B}T}+1)$, respectively.
      Here $E_{u,\bar{u}}$ is the energy and $\mu_{u}= -\mu_{\bar{u}}$ is the chemical potential of quarks. The Seebeck coefficient for QGP fluid can be investigated by following the same type of calculation as done in Section~(\ref{gf}), extended to three-dimensional phase space as,
     \begin{equation} 
     S_q =\frac{k_B}{Q_{u}}\left[  3 \frac{f_{3}(A)-f_{3}(A^{-1})}{f_{2}(A)+f_{2}(A^{-1})} - 2\mathfrak{h} \beta\right]~.\label{E33}
     \end{equation}
where $Q_u= (2/3)~e$ or $(-2/3)~ e$ is the charge of up quark (anti- quark). \\
 \section{Results and Discussion}
 \label{sec-Results}
  \begin{figure*}  
 	\centering 
 	\includegraphics[scale=0.55]{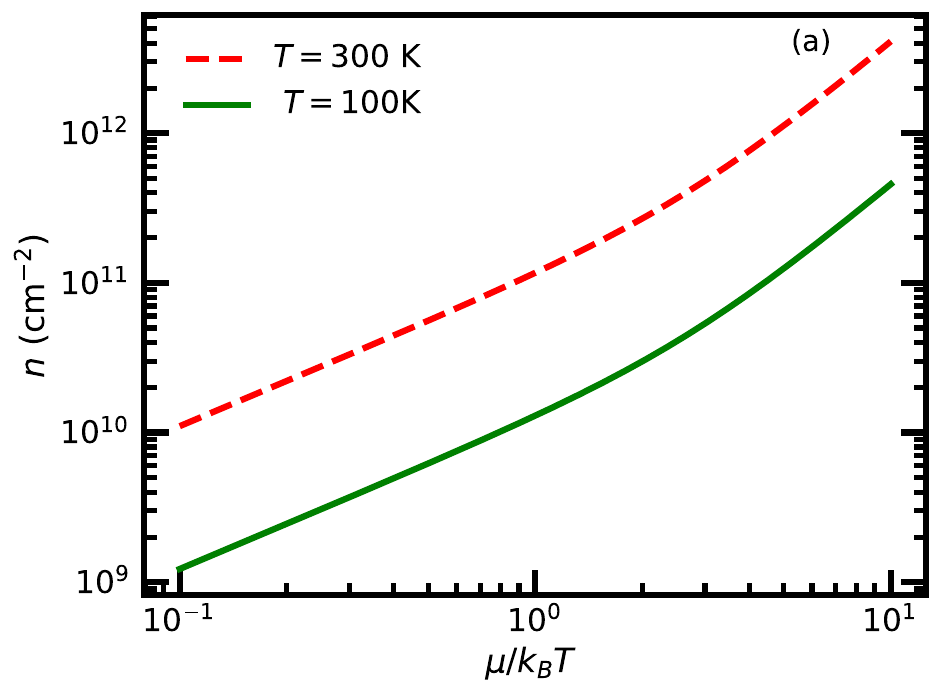}
 	\hfill
 	\includegraphics[scale=0.55]{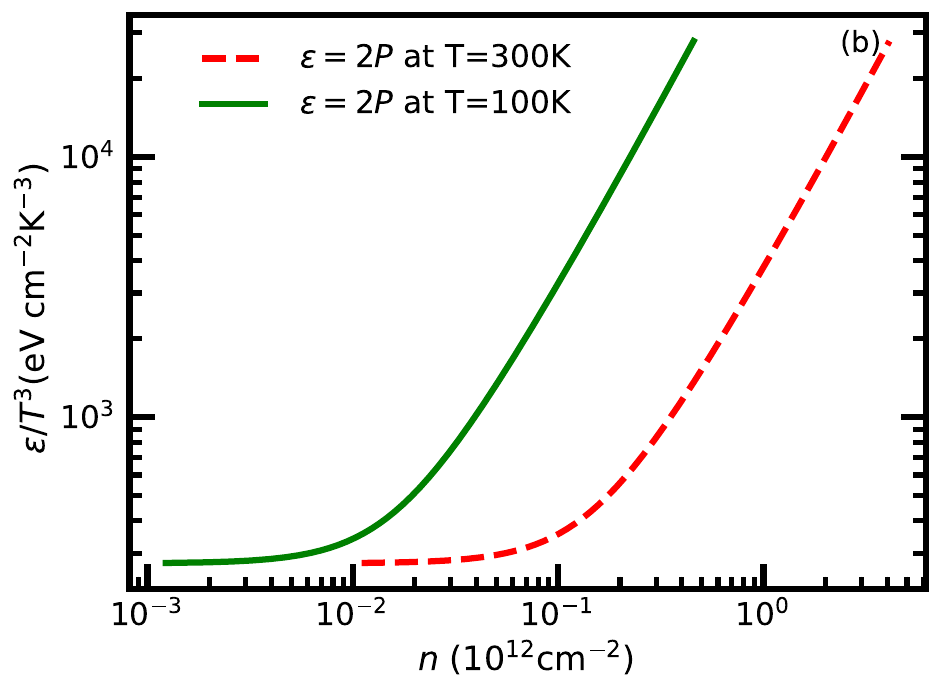}
 	\caption{(a): Net number density ($n$) as a function of scaled chemical potential ($\mu/k_{B} T$)  (b): Scaled energy density $\varepsilon/T^3$ as a function of $n$ at $T = 300$ and $100$~K for graphene.} 	\label{fig:1}
 \end{figure*}
In this section, we will explore the numerical results of our fluid based model of Seebeck coefficient for both graphene and QGP systems   .
 In Fig.~(\ref{fig:1}), we present all the fundamental thermodynamic quantities obtained from the kinetic theory formalism for the graphene system. Figure~(\ref{fig:1})~(a) shows the net carrier density as a function of normalized chemical potential at $T = 300$~K and $100$~K. As expected, following Fermi Dirac distribution, the carrier density of electrons dominates over the hole and increases exponentially with $\mu$. In Fig.~(\ref{fig:1})~(b), $\varepsilon$ or $P$ are plotted with net carrier density $(n)$. For graphene, both are related by $\varepsilon=2P$ with results represented for $T= 300$ K and $100$ K. 
  \begin{figure*}  
 	\centering 
 	\includegraphics[scale=0.53]{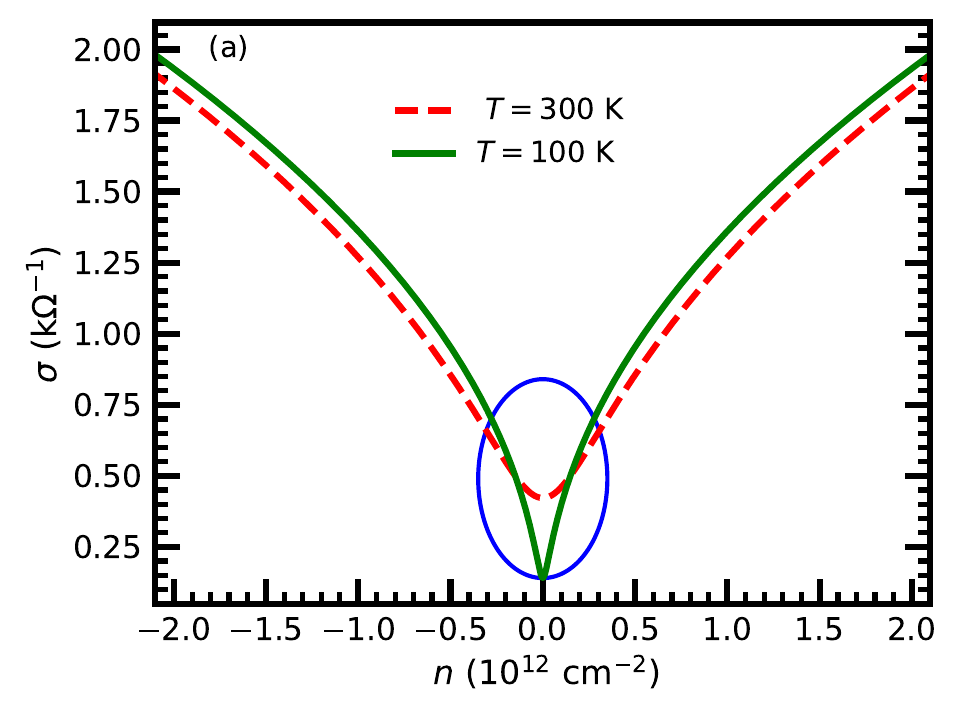}
 	\hfill
 	\includegraphics[scale=0.53]{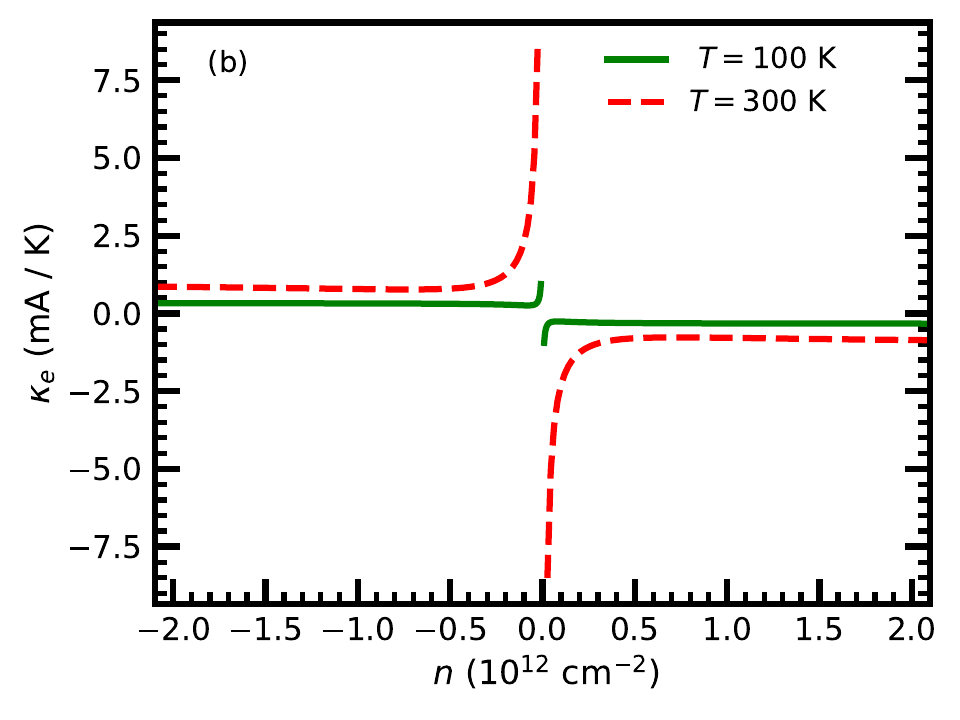}
 	\caption{Transport coefficients (a) electrical conductivity ($\sigma$) showing the quantum critical domain and (b) thermoelectric response coefficients ($k_e$) with reference to net carrier density ($n$) at $T = 300$ and $100$~K  for graphene.} 	\label{fig:2}
 \end{figure*}
Next, we generate the transport coefficient curves, which are linked with Seebeck coefficients for graphene. They are electrical charge conductivity $\sigma_{e}$ and the thermoelectric response coefficient $\kappa_{e}$.
In Fig.~(\ref{fig:2})~(a), $\sigma_{e}$ is plotted with $n$ at $T=300$ K and $T=100$ K. We consider the same relaxation time for electrons and hole as $\tau_R \approx 100$ fs~\cite{PhysRevLett.107.237401}. Electrical conductivity increases with net carrier density. From the expression of $\sigma_{e}$ Eq.~(\Ref{E22}), the contributions of electron and hole are added up due the square of charge ($q_i^2$). As a result, $\sigma_e$ is symmetric in the positive and negative sides of $n$. At the charge neutral point, conductivity is proportional to $T$, following the relation Eq.~(\ref{sigma0}),
\begin{equation}
	\sigma \Big|_{\mu=0} = 8\pi \ln{2} ~ \left[\frac{\tau_R e^2}{h^2} k_B T\right].
\end{equation}
As a result, the minima at $T = 100$~K are lower than those of $T = 300$~K. Around the quantum critical domain (at $\mu = 0$, $T \rightarrow 0$), one can expect that $\tau_R \propto \frac{\hbar}{k_B T}$~\cite{Lucasfong2018}. Therefore, electrical conductivity at the quantum critical point becomes a universal non-zero value up to a proportionality constant \cite{doi:10.1126/science.aat8687}. The blue circle marks the quantum critical conductivity domain. This marking highlights the distinction between the regions inside and outside the quantum critical domains, where the $T, \mu$ dependence of $\tau_R$ may differ. However, our attention in the present work is not towards the exact $\tau_R (T,\mu)$ or absolute $\sigma_e(T,\mu)$, so we have considered a simple constant $\tau_R$ with the relevant order of magnitude. Our primary interest lies in the Seebeck coefficient, where the exact $\tau_R (T,\mu)$ of two transport coefficients will cancel out. So, the graphs of the denominator $(\sigma_e)$ and the numerator $(\kappa_{e})$ of the Seebeck coefficient may be considered to understand their density dependence and overall order of magnitude, which will help to realize the exact values of the charge density dependence of $S_g$.  

In Fig.~(\ref{fig:2}) (b), we plotted the numerator of the Seebeck coefficient given in Eq.~(\ref{E23}). The quantity $\kappa_{e}$ is known as the thermoelectric response coefficient or charge flow per unit temperature. $\kappa_{e}$ is proportional to enthalpy per particle, which diverges (nonphysical) at vanishing charge carrier density. However, total enthalpy remains finite even at the charge neutral point \cite{Dwibedi2025,nayak2025electronhydrodynamicsgraphene}, which also can be understood from Fig.~(\ref{fig:1}) (b). From the expression of $\kappa_{e}$ in Eq.~(\ref{E23}), we see it is proportional to the electric charge $q_i$, which is negative for the electron and positive for the hole. As a result, the sign of $\kappa_{e}$ reversed from changing positive to negative $n$ or from an electron-dominated system to a hole-dominated counterpart.

  \begin{figure*}  
 	\centering 
 	\includegraphics[scale=0.53]{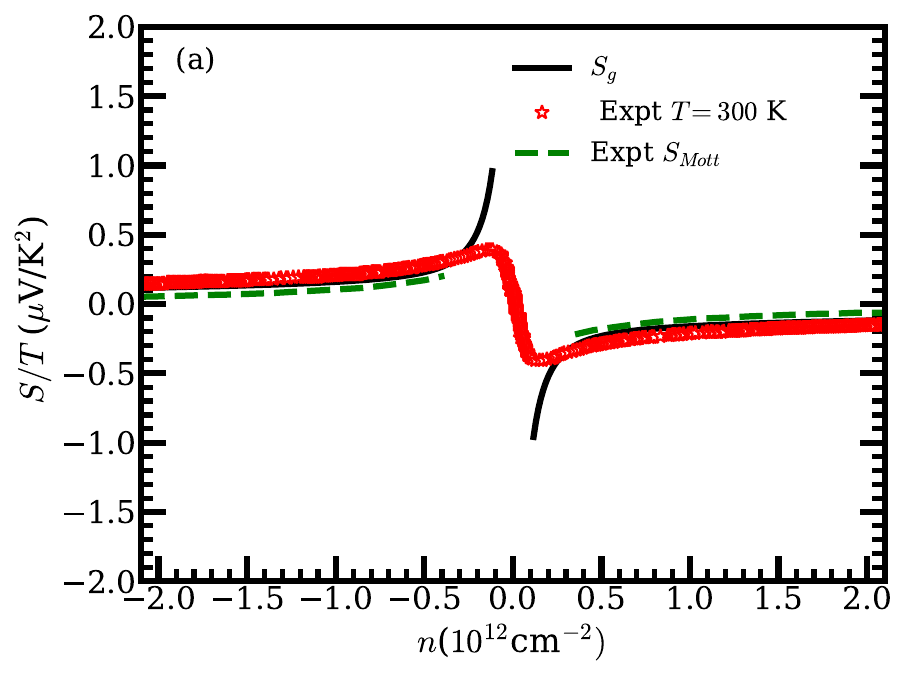}
 	\hfill
 	\includegraphics[scale=0.54]{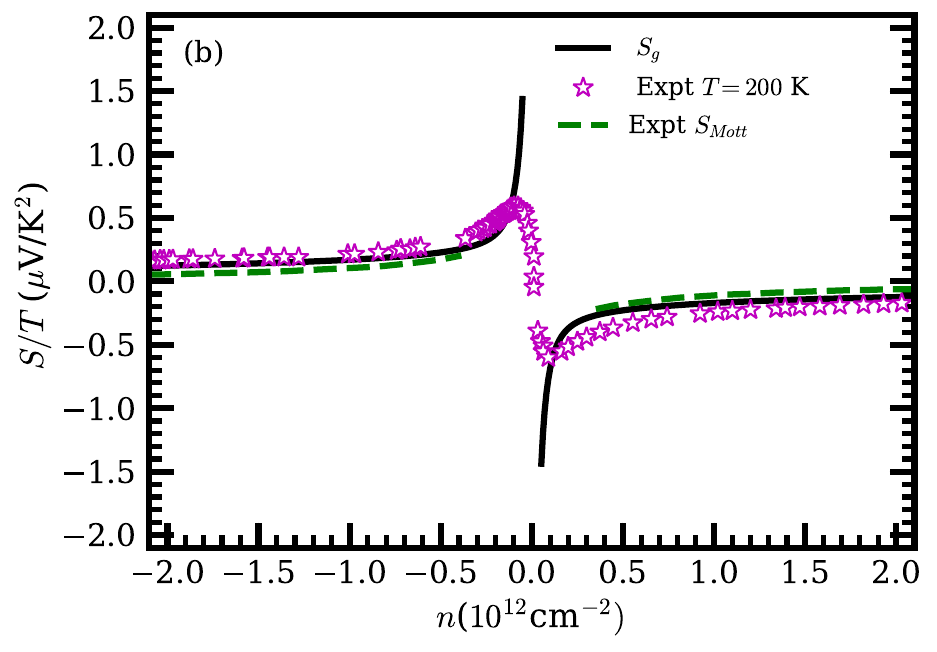}
 	\caption{ Scaled Seebeck coefficient $S/T$ as a function of net carrier density $n$ for graphene: (a) $T = 300$ K and  (b) $T = 200$ K. Black solid lines depict the present theoretical work $S_g$, while green dashed lines show the conventional Fermi liquid description of Mott relation (Eq.~(\ref{E10})) using  experimentally measured conductivity \cite{PhysRevLett.116.136802}. Experimental data are represented by red stars ($T=300$ K) and magenta stars ($T=200$ K) for comparison \cite{PhysRevLett.116.136802}.} 	\label{3}
 \end{figure*}
 
  \begin{figure*}  
 	\centering 
 	\includegraphics[scale=0.54]{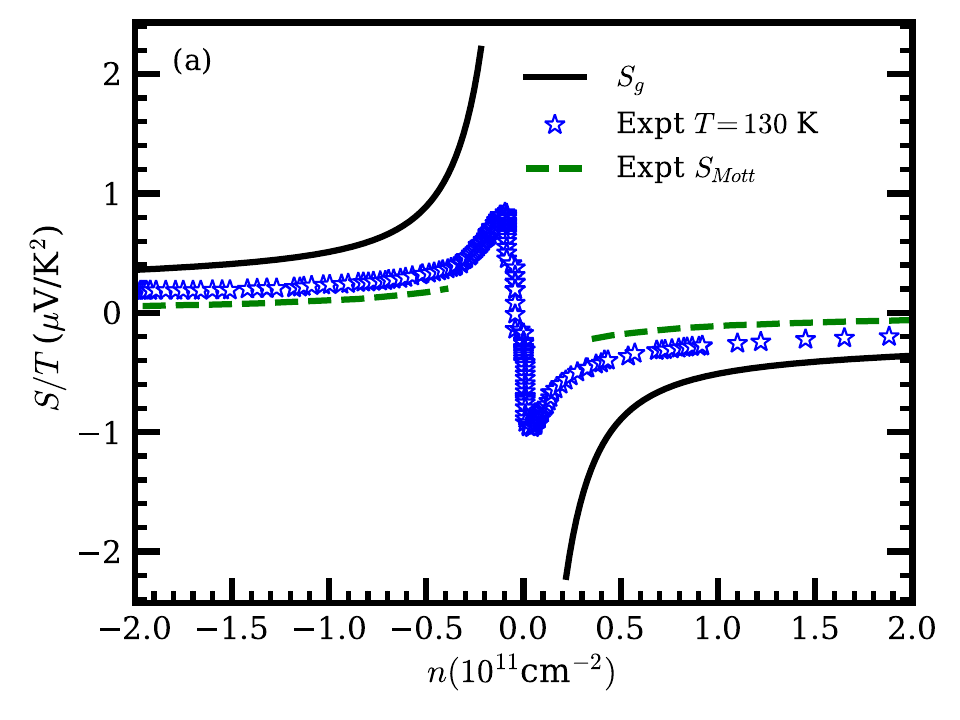}
 	\hfill
 	\includegraphics[scale=0.54]{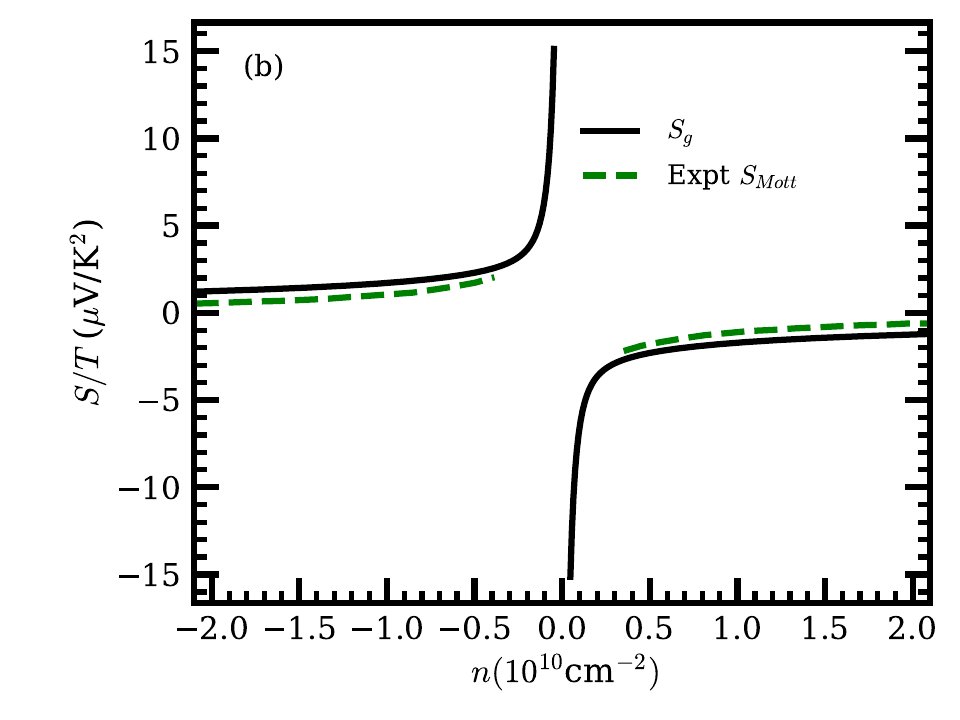}
 	\caption{Scaled Seebeck coefficient $S/T$ as a function of net carrier density $n$ for graphene: (a) $T = 130$ K and  (b) $T = 19$ K. Present theoretical work $S_g$ (black solid lines) are compared with the conventional Fermi liquid description of Mott relation (green dashed lines, Eq.~(\ref{E10})) and experimental data (blue stars) at ($T=130$ K) \cite{PhysRevLett.116.136802}. The theoretical predication of the $S_g$ is displayed in (b) at $T = 19$ K.  } 	\label{4}
 \end{figure*}
The Fig.~(\ref{3}) (a) presents the scaled Seebeck coefficient with net carrier density. The experimental data from Ref.~\cite{PhysRevLett.116.136802} of the Seebeck coefficient shows an enhanced value near the Dirac point at $T=300 $~K. 
 For comparison, we plot Mott's result, represented by green dashed-dotted line which is extracted from Ref.~\cite{PhysRevLett.116.136802}. Similar curves are plotted in Fig.~(\ref{3})~(b) and Fig.~(\ref{4})~(a) for $T=200$ K and $T=130$ K respectively. Reader can notice that experimental values are deviating from Mott's curve and the deviation is increasing as one goes from high to low (net) carrier density domain. The positive and negative density represent net electron $(n=n_e- n_h)$ and hole $(n= n_h- n_e)$ density domains. Now this deviation of experimental values of Seebeck coefficient from its traditional Mott values can be linked with the electron hydrodynamics property in graphene, which was recently observed experimentally \cite{gurzhi1963minimum,sulpizio2019visualizing,Bandurin2016,crossno2016observation,majumdar2025universality} and attempted to build an unconventional hydrodynamics theoretical framework \cite{Lucasfong2018,Narozhny2019uib,Narozhny:2022ncn,AnLucas2016,Dwibedi2025,nayak2025electronhydrodynamicsgraphene,win2024wied,win2025graphene,Aung:2023vrr,Aung:2025cbo}.
  In Fig~.(\ref{4})(b) we present the Seebeck coefficient for $T= 19$ K with the extracted data of Mott relation as reported in Ref.\cite{PhysRevLett.116.136802}. A deviation of the Seebeck coefficient from the Mott relation is observed near the charge neutrality point at very low temperature. This highlights significant violation of the Seebeck coefficient as compared to the Mott predication. Interestingly Ref.\cite{majumdar2025universality} has recently reported the substantial violation of the Wiedemann-Franz law in graphene for the same low temperature domain near the charge neutrality point due to the hydrodynamic effect. Consistent with this, for $T= 19$ K our result also shows a marked departure of the Seebeck coefficient from the Mott prediction. The violation is more prominent for low temperature domain, and serves as a clear evidence for the emergence of electron hydrodynamics.

 \begin{figure*}  
	\centering 
	\includegraphics[scale=0.55]{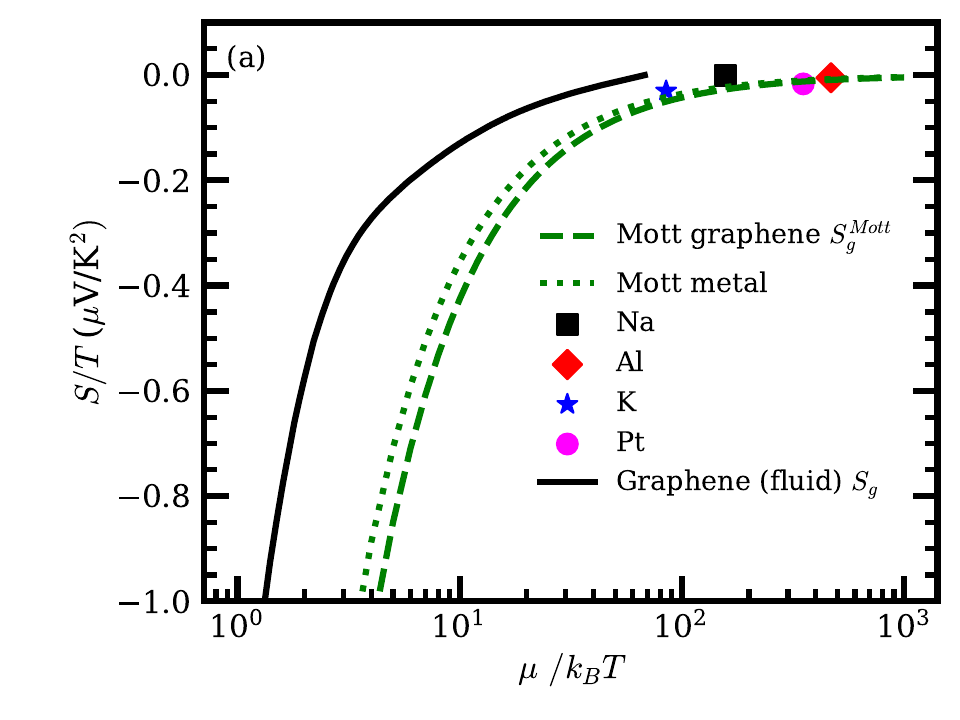}
	\hfill
	\includegraphics[scale=0.54]{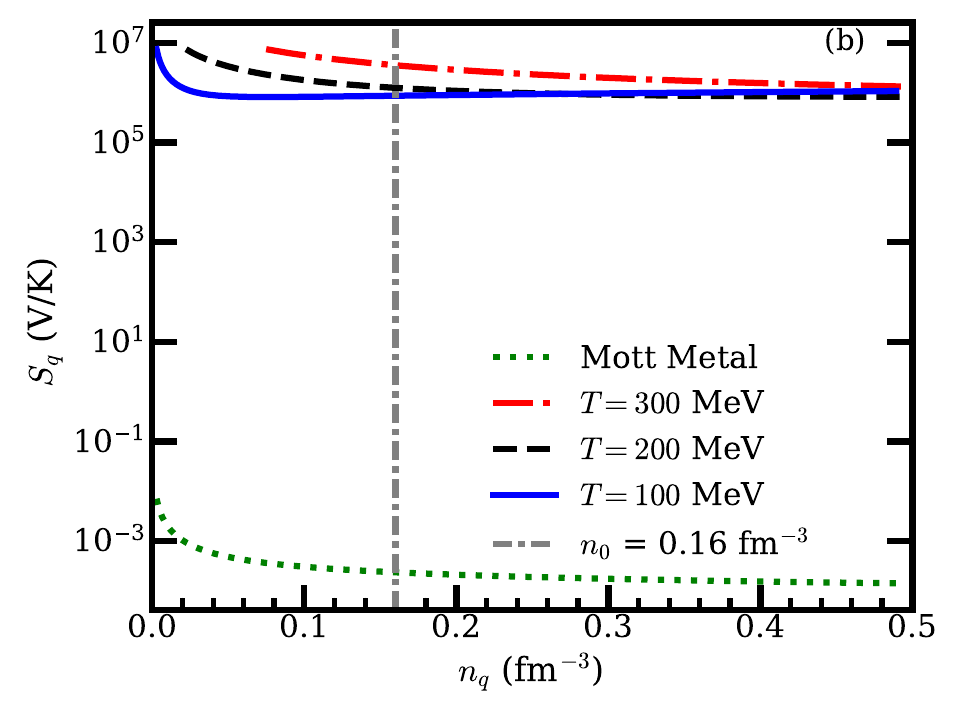}
	\caption{ (a) Scaled Seebeck coefficient $S/T$ vs reduced chemical potential $\mu/k_{B}T$ for graphene (fluid) $S_g$ and Mott relation (metal Eq.~(\ref{EMOTT_metal}) and graphene Eq.~(\ref{g_mott})) including some experimental data \cite{wikipedia_seebeck} for $T = 300$ K (b) Seebeck coefficient for QGP $S_q$ vs  net quark density $n_q$  over different temperatures in comparison with the predicted Mott relation for metal Eq.~(\ref{EMOTT_metal}). } 	\label{5}
\end{figure*}
Figure~(\ref{5}) (a) represents a comparative analysis of the scaled Seebeck coefficient (Eq.~(\ref{E32})) with respect to the $\mu/k_B T$, juxtaposed with the traditional Mott relation Eq.~(\ref{EMOTT_metal}). The conventional Mott relation for metals is shown by the green dotted line by considering the metal as a degenerate Fermi gas system by following  Eq.~(\ref{EMOTT_metal}). Here, for comparison, we have also estimated the Mott relation by assuming the degenerate gas limit for graphene with a linear dispersion relation $E=p v_F$. The Eq.(\ref{E10}) can be modified by considering $\frac{1}{\sigma(E)} \frac{\partial \sigma(E)}{\partial E}\Bigg|_{ \mu} =  \frac{3}{\mu}$ as follows,
\begin{equation}
	S^{Mott}_g = -\pi^2 \frac{k_B^{2}}{e} \frac{ T}{\mu}. \label{g_mott}
\end{equation}
The difference between Mott's relation of Seebeck coefficient for metal and graphene cases can be noticed in Fig.~(\ref{5}) (a) clearly, as well as can be understood by comparing Eqs.~(\ref{EMOTT_metal}) and (\ref{g_mott}) respectively.
Their different dispersion relations ($E \propto p^2, E \propto p$) are the main cause for this small difference. For hypothetical purpose we have extended our conventional Mott relation for metals in the entire domain of $\mu/k_B T$ but we know that Fermi energy of metals remain within $\mu=2-10$ eV range, Mott relation for metal well matched with the experimental values of Seebeck coefficients \cite{wikipedia_seebeck} which is represented in Fig.~(\ref{5}) (a) for standard reference.
In the DF domain $\mu/k_BT \ll 1$, we can clearly see that the fluid-based expression for graphene shows a deviation, whereas with the increasing value of $\mu$ all the results converge with each other in the FL domain ($\frac{\mu}{k_B T} >>1$). By compiling all the figures and equations, a non-fluid to fluid phase transition can be realized as follows. We may rename Eq.~(\ref{E32}) and Eq.~(\ref{g_mott}) as the fluid and non-fluid (Mott relation) of Seebeck coefficients in graphene. A non-fluid type Mott relation is expected in the FL domain, whereas the DF domain may favor a fluid description. So by tuning doping carrier density, when one transit from FL to DF domain, then a (Mott relation) non-fluid to fluid transition can be noticed in Seebeck coefficient, whose deviation from its Mott relation can be considered as a signature of fluid property.\\  
   In Fig.~(\ref{5})(b), we present the Seebeck coefficient ($S_{q}$) as a function of net quark density $(n_q)$ for the massless quark matter by following the Eq.~(\ref{E33}) at different temperatures $T=10^{12}$ K, $T=2\times10^{12}$ K and $T=3\times10^{12}$ K ($T=10^{12}$K $\approx 100$ MeV in natural units). For reference, we mark the nuclear saturation density $n_0=0.16$ fm$^{-3}$. For the high-density regime $(n_0 \ll n_q)$, the Seebeck coefficient for different values of temperature converges with each other. On the other hand, for the low-density regime $(n_q \ll n_0)$, a deviation from the Mott relation is observed. For comparison, the prediction from the Mott relation is also shown. Our results reveal a substantial deviation from the Mott relation. In terms of order of magnitude, the Seebeck coefficient of QGP fluid is $10^8- 10^{10}$ times larger than its Mott value, which is valid for a non-fluid picture of QGP. Earlier studies~\cite{Abhishek:2020wjm,PhysRevD.103.054024,PhysRevD.111.096011,PhysRevD.109.014018,Das2021,PhysRevD.104.076021} also report that the order of magnitude of the Seebeck coefficient in QGP shows a substantial departure from Mott's estimation. Direct measurement of the Seebeck coefficient may not be possible, like other transport coefficients and their ratios, but an indirect measurement via dilepton and photon rates may be possible. One of our future research is planned in this direction.  
  \section{Summary and conclusions}\label{sum}
 In this work, we studied the thermoelectric properties of graphene and QGP using the kinetic theory formalism in the hydrodynamic regime. To obtain the Seebeck coefficient, we solved the Boltzmann transport equation under the relaxation time approximation. The transport properties of graphene are governed by massless charge carriers, electrons and holes, where they follow the linear dispersion relation. Quarks and anti-quarks in quark-gluon plasma exhibit analogous transport properties due to their linear dispersion relations. Mathematically, the thermoelectric transport coefficients and their ratios, such as the Seebeck coefficient for graphene and quark-gluon plasma, share similar structures in terms of Fermi integral functions but differ due to the different dimensionality of these two systems. At high-charge-carrier densities, corresponding to the Fermi liquid domain, the Mott relation of Seebeck coefficient values is expected. This is consistent with the standard agreement between experimental values and the Mott theory for non-relativistic electrons in conventional metals. Calibrating the Mott theory with experiments in the metallic Fermi energy range $\mu=2-10$ eV, we have plotted the curves of the Mott and hydrodynamic theories for graphene. Interestingly, both merge in a high charge carrier density regime or a metallic Fermi energy regime, whereas they become well separated in a low charge carrier density regime or a Dirac fluid regime. Experimental measurement of the Seebeck coefficient of graphene favors towards the breakdown of the conventional Mott relation and fluid-based Seebeck response. An important outcome of the present analysis is that the violation of the Mott relation emerges naturally from the fluid-based treatment of the couple charge and thermal currents, with the enthalpy per particle playing a crucial role. In the QGP sector as well, one can expect this fluid behavior and a (strong) violation of the Mott relation in the Seebeck coefficient.
 
\section{Acknowledgement}
SN acknowledges Ministry of Education, Govt. of India.
JD acknowledges financial support from JINR Postdoctoral Programme.
SG acknowledges  the Board of Research in Nuclear Sciences and Department of Atomic Energy, Govt. of India, under Grant No. 57/14/01/2024-BRNS/313. 
The authors are also grateful 
to Thandar Zaw Win, Ashutosh Diwedi, Dani Rose J Marratukalam, and Ankit Panda  for fruitful discussions.

\section{appendix}
\subsection{Calculation of $\delta f$} \label{ape1}
To calculate $\delta f_i$ in Eq.~(\ref{E14}) we have used Boltzmann transport equation in RTA framework. It can be expressed as follows,
\begin{equation}
	\frac{\partial f_{i}}{\partial t} + \vec{v}_{F} \cdot \frac{\partial f_{i}}{\partial \vec{x}} + q_{i} \vec{E} \cdot \frac{\partial f_{i}}{\partial \vec{p}} = -\frac{\delta f_{i}}{\tau_i}  ~,\label{a1}
\end{equation}
where $\tau_{i}$ is the relaxation time of the particle. The Eq.~(\ref{a1}) can be solved by using the Eq.~(\ref{E17}) and (\ref{E18}). Afterwards, the values of  $\alpha_{1}$ and $\alpha_{2}$
can be determined.
The first term in the L.H.S of Eq.~(\ref{a1}) can be calculated as follows,
\begin{equation}
	\frac{\partial f_{i}}{\partial t} = \frac{\partial f^{0}_{i}}{\partial t} +\frac{\partial \delta f_{i}}{\partial t} ~. \label{a2}
\end{equation}
Again the first term on the R.H.S of Eq.~(\ref{a2}) can be written as,
\begin{eqnarray}
		\frac{\partial f^{0}_{i}}{\partial t} &=& \frac{\partial }{\partial t} \left[\frac{1}{\exp(\frac{E-Q_i\mu(\vec{x})}{ k_B T(\vec{x},t)})+1}\right]\nn\\
		&=&- \left(\frac{E-Q_i\mu(\vec{x})}{ T(\vec{x},t)}\right) \dot{T} \frac{\partial f^{0}_{i}}{\partial E} ~. \label{a3}
\end{eqnarray}
The $2^{nd}$ term on the R.H.S of Eq.~(\ref{a2}) using Eq.~(\ref{E17}) and (\ref{E18}) can be calculated as follows,
\begin{eqnarray}
	\frac{\partial \delta f_{i}}{\partial t} &=& \frac{\partial }{\partial t} [\vec{v}_F.(\alpha_{1} \vec{E} + \alpha_{2} \vec{\nabla T})] \frac{\partial f^{0}_{i}}{\partial E} \nn\\
	&=&[\vec{v}_F . (\dot{\alpha_{1}}\vec{E}+ \alpha_{1} \dot{\vec{E}}+\dot{\alpha_{2}}\vec{E}+\alpha_{2} \dot{\vec{E}} )]~\frac{\partial f^{0}_{i}}{\partial E} ~. \label{a4}
\end{eqnarray}
Eq.~(\ref{a2}) now take the form,
\begin{equation}
	\begin{split}
		\frac{\partial f _{i}}{\partial t} = \Bigg[
		&-\left(\frac{E-Q_i\mu(\vec{x})}{k_B T(\vec{x},t)}\right) \dot{T} \\
		&+ \vec{p} \cdot \big(
		\dot{\alpha_{1}}\vec{E} + \alpha_{1} \dot{\vec{E}}
		+ \dot{\alpha_{2}}\vec{E} + \alpha_{2} \vec{\nabla \dot{T}}
		\big)
		\Bigg]
		\frac{\partial f^{0}_{i}}{\partial E} ~.
	\end{split}
	~\label{a5}
\end{equation}
Now considering the second term in the Eq.~(\ref{a1}),
\begin{eqnarray}
	\frac{\partial  f_{i}}{\partial \vec{x}} &=& \frac{\partial f^{0}_{i}}{\partial \vec{x}}  +\frac{\partial  (\delta f_{i})}{\partial \vec{x}} \nn\\
	&=& \frac{\partial }{\partial \vec{x}} \left[\frac{1}{\exp(\frac{E-Q_i\mu(\vec{x})}{k_B T(\vec{x},t)})+1}\right] +0 \nn\\
	&=& -\frac{\exp\left(\frac{E-Q_i\mu(\vec{x})}{k_B T(\vec{x},t)}\right)}{\{\exp(\frac{E-Q_i\mu(\vec{x})}{k_B T(\vec{x},t)})+1\}^{2}}\, \left[\frac{\partial }{\partial \vec{x}}\left(\frac{E-Q_i\mu(\vec{x})}{{k_B T(\vec{x},t)}}\right)\right]\nn\\
	&=& -\left(\frac{E- Q_i\mathfrak{h} }{T}\right) ~\vec{\nabla} T~ \frac{\partial f^{0}_{i}}{\partial E}~. \label{a6}
\end{eqnarray}	  
In the last line we used the Gibb's–Duhem relation
\begin{equation}
	\frac{T}{\mathfrak{h}} \frac{\partial (\mu/T)}{\partial x} = \frac{1}{n\mathfrak{h}} \frac{\partial P}{\partial x} - \frac{1}{T} \, \frac{\partial T}{\partial x}, 
\end{equation}
where $\mathfrak{h} = \frac{E+P}{n}$ is the enthalpy per particle. Moreover, we considered the local rest frame (LRF) of fluid with $\frac{\partial P}{\partial \vec{x}} = 0$.\\
The third term in the LHS of Eq.(\ref{a1}) can be calculated using Eq.~(\ref{E17}) as follows,
\begin{eqnarray}
	q_i \vec{E} \cdot \frac{\partial f}{\partial \vec{p}} &=& q_i \vec{E} \cdot \frac{\partial f^{0}_{i}}{\partial \vec{p}} + q_i\vec{E} \cdot \frac{\partial (\delta f_{i})}{\partial \vec{p}}\nn\\
	\implies q_i\vec{E} \, \frac{\partial f}{\partial p} &=& q_i \vec{E} \cdot \vec{v}_F \frac{\partial f^{0}_{i}}{\partial E} + 0,
	\label{a7}
\end{eqnarray}
where in the last line we neglected the second order contribution in the distribution function, on the right hand side.
Now Eq.(\ref{a1}) can be written as,
\begin{align}
	& - \frac{(E-Q_i\mu)}{T} \dot{T} + \vec{v}_F \cdot (\dot{\alpha_{1}}\vec{E} + \alpha_{1} \dot{\vec{E}}+\dot{\alpha_{2}}\vec{E}+\alpha_{2} \vec{\nabla} \dot{T})
	\nn\\
	& -\left(\frac{E- Q_i\mathfrak{h} }{T}\right) (\vec{v}_F \cdot \vec{\nabla} T)~  
	+ q_i ~ \vec{v}_{F} \cdot \vec{E}\nn\\
	&= -\frac{\vec{v}_{F}} {\tau_{i}} \cdot (\alpha_{1} \vec{E} + \alpha_{2} \vec{\nabla} T). \label{a8}
\end{align}
Comparing the coefficient of $v_{F} . \vec{E}$ on both sides,
\begin{equation}
	\dot{\alpha_{1}}  + q_i = - \frac{\alpha _{1}}{\tau_{i}}.  \label{a9}
\end{equation}
 By solving,
\begin{equation}
	\dot{\alpha_{1}}  = -\left( \frac{\alpha _{1}}{\tau_{i}} + q_i\right).  \label{a10}
\end{equation}  
Similarly, comparing the coefficient of   $\vec{v}_F . \vec{\nabla T}$,        
\begin{equation}
	\dot{\alpha_{2}}  = -\left( \frac{\alpha _{2}}{\tau_{i}} -\frac{E-Q_i\mathfrak{h}}{T }\right). \label{a11} 
\end{equation}  
After solving Eq.~(\ref{a10}) and (\ref{a11}),
\begin{equation}
	\alpha_{1}  =  - q_i ~ \tau_{i}  ~,~~ \alpha_{2}=\frac{E-Q_i \mathfrak{h}}{T} \tau_{i}~. \label{a12}
\end{equation}
Using Eq.~(\ref{a12}) in (\ref{E19}) the following result is obtained,
\begin{equation}
	\delta f_{i} =  \left[- q_{i} \tau_i ( \vec{v}_F . \vec{E})  + \left(\frac{E- Q_i\mathfrak{h} }{T}\right)\tau_i (\vec{v}_{F} .\vec{\nabla} T )\right] \frac{\partial f^{0}_{i}}{\partial E} ~. \label{a13}
\end{equation}

\subsection{Details of hydrodynamic formulation of charge carriers in graphene} \label{ape2}
In this subsection we have calculated the Seebeck coefficient for graphene by considering the fluidic nature of electron  with its linear dispersion relation. \\
Using Eq.~(\ref{a13}) in Eq.~(\ref{E14}) the current density takes the following form,
\begin{equation}
	\vec{j}= \sum_{i} \frac{q_i g}{2} \int \frac{d^{2}p}{h^{2}}~ \vec{v}_{F}^{2} ~\tau_i \left[- q_{i}   \vec{E}  +\left(\frac{E- Q_i\mathfrak{h} }{T}\right) \vec{\nabla} T \right] \frac{\partial f^{0}_{i}}{\partial E} ~. \label{a14}
\end{equation}
Under equilibrium condition there will be no electric current flows through the graphene sample with  temperature gradient,
$i.e.,$ $\vec{j}=0$. \\
Eq.~(\ref{a14}) becomes 
\begin{eqnarray}
   &&-\sum_{i} \frac{q_i^{2} g}{2} \int \frac{d^{2}p}{h^{2}}~ \vec{v}_{F}^{2} ~\tau_i   \vec{E}~ \frac{\partial f^{0}_{i}}{\partial E} \nn\\
  = \sum_{i} \frac{q_i g}{2} 
   &&\int \frac{d^{2}p}{h^{2}}~ \vec{v}_{F}^{2} ~\tau_i ~\left(\frac{E- Q_i\mathfrak{h} }{T}\right) ~ \vec{\nabla} T~ \frac{\partial f^{0}_{i}}{\partial E}. \label{a15}
\end{eqnarray}
According to the semi-classical theory, the current density and heat current density can be written in a following form,
\begin{eqnarray}
	j &=& \sigma_{e} \vec{E}- \kappa_{e} (\vec{\nabla}T)=-\sigma_{e} \vec{\nabla} V- \kappa_{e} (\vec{\nabla}T) ~,\label{a16a} \\
	j^{q}&=& \sigma_{th} \vec{E}+ \kappa_{th} (-\vec{\nabla}T) ~.\label{a16}
\end{eqnarray}
If $\vec{j}=0$ then,
\begin{equation}
 \sigma_{e} \vec{E}= \kappa_{e} (\vec{\nabla}T)~. \label{a17}
\end{equation}
The Seebeck coefficient from Eq.~(\ref{a17}) can be written as,
\begin{equation}
	S_g= \frac{\vec{E}}{\vec{\nabla} T}= \frac{\kappa_{e}}{\sigma_{e}}. \label{a17a}
\end{equation}
\\
Comparing the Eq.~(\ref{a14}) with Eq.~(\ref{a16a}), we get the integral form of the electrical conductivity as follows,
\begin{equation}
	\sigma_e=-\sum_{i} \frac{q_i^{2} g}{2} \int \frac{d^{2}p}{h^{2}}~ \vec{v}_{F}^{2} ~\tau_i~ \frac{\partial f^{0}_{i}}{\partial E}. \label{a18}
\end{equation}
To derive step-by-step the expression for electrical conductivity given in  Eq.~(\ref{a18}), we proceed under the assumption of a common relaxation time, $\tau_i= \tau_R$ as follows,
\bea
\sigma_e 
&=& - \frac{e^{2} ~ \tau_{R} ~ g ~ v_{F}^{2}}{2} 
\Bigg[ 
\int \frac{d^{2}p}{h^{2}} ~ \frac{\partial f^{0}_{e}}{\partial E} 
+ \int \frac{d^{2}p}{h^{2}} ~ \frac{\partial f^{0}_{h}}{\partial E} 
\Bigg] \nn\\
&=&  \frac{e^{2} ~ \tau_{R} ~ g ~ \pi v_{F}^{2}}{h^{2}} 
\Bigg[ 
\int p ~ dp ~ \frac{\partial f^{0}_{e}}{\partial \mu} 
+ \int p ~ dp ~ \frac{\partial f^{0}_{h}}{\partial \mu} 
\Bigg].
\eea
Here we have used $\frac{\partial f^0_{e,h}}{\partial E}=- \frac{\partial f^0_{e,h} }{\partial \mu} = \beta f^0_{e,h} (1-f^0_{e,h})$, where $f^0_{e,h}= \frac{1}{exp ~\beta(E\mp\mu)\pm 1}$. Now by using the Fermi integral provided by Eq~. (\ref{E28}) and the identity $\frac{\partial f_\nu (A)}{\partial (\frac{\mu}{k_B T})}= f_{\nu-1} (A)$ the above equation can be written as,
\begin{equation}
	\sigma_e = \frac{ e^{2}g ~\pi~ \tau_{R}~ k_{B} ~T}{h^2} ~\{f_{1}(A)+f_{1}(A^{-1}) \}~.\label{a20}
\end{equation}
Similarly by comparing the Eq.~(\ref{a14}) and  Eq.~(\ref{a16a}), we can get the thermoelectric response coefficient as follows, 
\begin{equation}
	k_{e}=-\sum_{i} \frac{q_i g}{2} 
	\int \frac{d^{2}p}{h^{2}}~ \vec{v}_{F}^{2} ~\tau_i ~\left(\frac{E- Q_i\mathfrak{h} }{T}\right) ~ \vec{\nabla }T~ \frac{\partial f^{0}_{i}}{\partial E}.\label{a19}
\end{equation}
Moving to the calculation of $\kappa_e$ from Eq.~(\ref{a19}) we get the following expression,
\begin{align}
	k_{e} &=  \frac{e\, g \,\pi\, \tau_{i} \,k^{2}_{B}\,T}{h^2}  
	\Big[ 2 \{f_{2}(A)-f_{2}(A^{-1})\} \nn\\
	&\quad - \mathfrak{h} \beta \,\{f_{1}(A)+f_{1}(A^{-1})\}\Big] .
	\label{a21}
\end{align}
From Eqs.~(\ref{a20}) and (\ref{a21}) the Seebeck coefficient for graphene system is found as follows,
\begin{equation}
	S_g =\frac{k_B}{e}\left[  2 \frac{f_{2}(A)-f_{2}(A^{-1})}{f_{1}(A)+f_{1}(A^{-1})} - \mathfrak{h} \beta\right].\label{a23}
\end{equation}

\bibliography{ref}

\begin{thebibliography}{79}%
\makeatletter
\providecommand \@ifxundefined [1]{%
 \@ifx{#1\undefined}
}%
\providecommand \@ifnum [1]{%
 \ifnum #1\expandafter \@firstoftwo
 \else \expandafter \@secondoftwo
 \fi
}%
\providecommand \@ifx [1]{%
 \ifx #1\expandafter \@firstoftwo
 \else \expandafter \@secondoftwo
 \fi
}%
\providecommand \natexlab [1]{#1}%
\providecommand \enquote  [1]{``#1''}%
\providecommand \bibnamefont  [1]{#1}%
\providecommand \bibfnamefont [1]{#1}%
\providecommand \citenamefont [1]{#1}%
\providecommand \href@noop [0]{\@secondoftwo}%
\providecommand \href [0]{\begingroup \@sanitize@url \@href}%
\providecommand \@href[1]{\@@startlink{#1}\@@href}%
\providecommand \@@href[1]{\endgroup#1\@@endlink}%
\providecommand \@sanitize@url [0]{\catcode `\\12\catcode `\$12\catcode
  `\&12\catcode `\#12\catcode `\^12\catcode `\_12\catcode `\%12\relax}%
\providecommand \@@startlink[1]{}%
\providecommand \@@endlink[0]{}%
\providecommand \url  [0]{\begingroup\@sanitize@url \@url }%
\providecommand \@url [1]{\endgroup\@href {#1}{\urlprefix }}%
\providecommand \urlprefix  [0]{URL }%
\providecommand \Eprint [0]{\href }%
\providecommand \doibase [0]{https://doi.org/}%
\providecommand \selectlanguage [0]{\@gobble}%
\providecommand \bibinfo  [0]{\@secondoftwo}%
\providecommand \bibfield  [0]{\@secondoftwo}%
\providecommand \translation [1]{[#1]}%
\providecommand \BibitemOpen [0]{}%
\providecommand \bibitemStop [0]{}%
\providecommand \bibitemNoStop [0]{.\EOS\space}%
\providecommand \EOS [0]{\spacefactor3000\relax}%
\providecommand \BibitemShut  [1]{\csname bibitem#1\endcsname}%
\let\auto@bib@innerbib\@empty
\bibitem [{\citenamefont {Cao}\ \emph {et~al.}(2011)\citenamefont {Cao},
  \citenamefont {Elliott}, \citenamefont {Joseph}, \citenamefont {Wu},
  \citenamefont {Petricka}, \citenamefont {Sch\"afer},\ and\ \citenamefont
  {Thomas}}]{Cao:2010wa}%
  \BibitemOpen
  \bibfield  {author} {\bibinfo {author} {\bibfnamefont {C.}~\bibnamefont
  {Cao}}, \bibinfo {author} {\bibfnamefont {E.}~\bibnamefont {Elliott}},
  \bibinfo {author} {\bibfnamefont {J.}~\bibnamefont {Joseph}}, \bibinfo
  {author} {\bibfnamefont {H.}~\bibnamefont {Wu}}, \bibinfo {author}
  {\bibfnamefont {J.}~\bibnamefont {Petricka}}, \bibinfo {author}
  {\bibfnamefont {T.}~\bibnamefont {Sch\"afer}},\ and\ \bibinfo {author}
  {\bibfnamefont {J.~E.}\ \bibnamefont {Thomas}},\ }\bibfield  {title}
  {\bibinfo {title} {{Universal Quantum Viscosity in a Unitary Fermi Gas}},\
  }\href {https://doi.org/10.1126/science.1195219} {\bibfield  {journal}
  {\bibinfo  {journal} {Science}\ }\textbf {\bibinfo {volume} {331}},\ \bibinfo
  {pages} {58} (\bibinfo {year} {2011})},\ \Eprint
  {https://arxiv.org/abs/1007.2625} {arXiv:1007.2625 [cond-mat.quant-gas]}
  \BibitemShut {NoStop}%
\bibitem [{\citenamefont {Bernhard}\ \emph {et~al.}(2019)\citenamefont
  {Bernhard}, \citenamefont {Moreland},\ and\ \citenamefont
  {Bass}}]{Bernhard2019}%
  \BibitemOpen
  \bibfield  {author} {\bibinfo {author} {\bibfnamefont {J.~E.}\ \bibnamefont
  {Bernhard}}, \bibinfo {author} {\bibfnamefont {J.~S.}\ \bibnamefont
  {Moreland}},\ and\ \bibinfo {author} {\bibfnamefont {S.~A.}\ \bibnamefont
  {Bass}},\ }\bibfield  {title} {\bibinfo {title} {Bayesian estimation of the
  specific shear and bulk viscosity of quark--gluon plasma},\ }\href
  {https://doi.org/10.1038/s41567-019-0611-8} {\bibfield  {journal} {\bibinfo
  {journal} {Nature Physics}\ }\textbf {\bibinfo {volume} {15}},\ \bibinfo
  {pages} {1113} (\bibinfo {year} {2019})}\BibitemShut {NoStop}%
\bibitem [{\citenamefont {Schenke}\ \emph {et~al.}(2012)\citenamefont
  {Schenke}, \citenamefont {Tribedy},\ and\ \citenamefont
  {Venugopalan}}]{PhysRevLett.108.252301}%
  \BibitemOpen
  \bibfield  {author} {\bibinfo {author} {\bibfnamefont {B.}~\bibnamefont
  {Schenke}}, \bibinfo {author} {\bibfnamefont {P.}~\bibnamefont {Tribedy}},\
  and\ \bibinfo {author} {\bibfnamefont {R.}~\bibnamefont {Venugopalan}},\
  }\bibfield  {title} {\bibinfo {title} {Fluctuating glasma initial conditions
  and flow in heavy ion collisions},\ }\href
  {https://doi.org/10.1103/PhysRevLett.108.252301} {\bibfield  {journal}
  {\bibinfo  {journal} {Phys. Rev. Lett.}\ }\textbf {\bibinfo {volume} {108}},\
  \bibinfo {pages} {252301} (\bibinfo {year} {2012})}\BibitemShut {NoStop}%
\bibitem [{\citenamefont {Garbiso}\ and\ \citenamefont
  {Kaminski}(2020)}]{Garbiso2020}%
  \BibitemOpen
  \bibfield  {author} {\bibinfo {author} {\bibfnamefont {M.}~\bibnamefont
  {Garbiso}}\ and\ \bibinfo {author} {\bibfnamefont {M.}~\bibnamefont
  {Kaminski}},\ }\bibfield  {title} {\bibinfo {title} {Hydrodynamics of simply
  spinning black holes {\&} hydrodynamics for spinning quantum fluids},\ }\href
  {https://doi.org/10.1007/JHEP12(2020)112} {\bibfield  {journal} {\bibinfo
  {journal} {Journal of High Energy Physics}\ }\textbf {\bibinfo {volume}
  {2020}},\ \bibinfo {pages} {112} (\bibinfo {year} {2020})}\BibitemShut
  {NoStop}%
\bibitem [{\citenamefont {Del Zanna}\ \emph {et~al.}(2013)\citenamefont
  {Del Zanna}, \citenamefont {Chandra}, \citenamefont {Inghirami},
  \citenamefont {Rolando}, \citenamefont {Beraudo}, \citenamefont {De Pace},
  \citenamefont {Pagliara}, \citenamefont {Drago},\ and\ \citenamefont
  {Becattini}}]{DelZaznna2013}%
  \BibitemOpen
  \bibfield  {author} {\bibinfo {author} {\bibfnamefont {L.}~\bibnamefont
  {Del Zanna}}, \bibinfo {author} {\bibfnamefont {V.}~\bibnamefont {Chandra}},
  \bibinfo {author} {\bibfnamefont {G.}~\bibnamefont {Inghirami}}, \bibinfo
  {author} {\bibfnamefont {V.}~\bibnamefont {Rolando}}, \bibinfo {author}
  {\bibfnamefont {A.}~\bibnamefont {Beraudo}}, \bibinfo {author} {\bibfnamefont
  {A.}~\bibnamefont {De Pace}}, \bibinfo {author} {\bibfnamefont
  {G.}~\bibnamefont {Pagliara}}, \bibinfo {author} {\bibfnamefont
  {A.}~\bibnamefont {Drago}},\ and\ \bibinfo {author} {\bibfnamefont
  {F.}~\bibnamefont {Becattini}},\ }\bibfield  {title} {\bibinfo {title}
  {Relativistic viscous hydrodynamics for heavy-ion collisions with echo-qgp},\
  }\href {https://doi.org/10.1140/epjc/s10052-013-2524-5} {\bibfield  {journal}
  {\bibinfo  {journal} {The European Physical Journal C}\ }\textbf {\bibinfo
  {volume} {73}},\ \bibinfo {pages} {2524} (\bibinfo {year}
  {2013})}\BibitemShut {NoStop}%
\bibitem [{\citenamefont {Schenke}\ \emph {et~al.}(2010)\citenamefont
  {Schenke}, \citenamefont {Jeon},\ and\ \citenamefont
  {Gale}}]{PhysRevC.82.014903}%
  \BibitemOpen
  \bibfield  {author} {\bibinfo {author} {\bibfnamefont {B.}~\bibnamefont
  {Schenke}}, \bibinfo {author} {\bibfnamefont {S.}~\bibnamefont {Jeon}},\ and\
  \bibinfo {author} {\bibfnamefont {C.}~\bibnamefont {Gale}},\ }\bibfield
  {title} {\bibinfo {title} {(3+1)d hydrodynamic simulation of relativistic
  heavy-ion collisions},\ }\href {https://doi.org/10.1103/PhysRevC.82.014903}
  {\bibfield  {journal} {\bibinfo  {journal} {Phys. Rev. C}\ }\textbf {\bibinfo
  {volume} {82}},\ \bibinfo {pages} {014903} (\bibinfo {year}
  {2010})}\BibitemShut {NoStop}%
\bibitem [{\citenamefont {Jaiswal}\ and\ \citenamefont
  {Roy}(2016)}]{jaiswal_AHEP}%
  \BibitemOpen
  \bibfield  {author} {\bibinfo {author} {\bibfnamefont {A.}~\bibnamefont
  {Jaiswal}}\ and\ \bibinfo {author} {\bibfnamefont {V.}~\bibnamefont {Roy}},\
  }\bibfield  {title} {\bibinfo {title} {Relativistic hydrodynamics in
  heavy-ion collisions: General aspects and recent developments},\ }\href
  {https://doi.org/https://doi.org/10.1155/2016/9623034} {\bibfield  {journal}
  {\bibinfo  {journal} {Advances in High Energy Physics}\ }\textbf {\bibinfo
  {volume} {2016}},\ \bibinfo {pages} {9623034} (\bibinfo {year} {2016})},\
  \Eprint
  {https://arxiv.org/abs/https://onlinelibrary.wiley.com/doi/pdf/10.1155/2016/9623034}
  {https://onlinelibrary.wiley.com/doi/pdf/10.1155/2016/9623034} \BibitemShut
  {NoStop}%
\bibitem [{\citenamefont {Ding}\ \emph {et~al.}(2016)\citenamefont {Ding},
  \citenamefont {Kaczmarek},\ and\ \citenamefont {Meyer}}]{PhysRevD.94.034504}%
  \BibitemOpen
  \bibfield  {author} {\bibinfo {author} {\bibfnamefont {H.-T.}\ \bibnamefont
  {Ding}}, \bibinfo {author} {\bibfnamefont {O.}~\bibnamefont {Kaczmarek}},\
  and\ \bibinfo {author} {\bibfnamefont {F.}~\bibnamefont {Meyer}},\ }\bibfield
   {title} {\bibinfo {title} {Thermal dilepton rates and electrical
  conductivity of the qgp from the lattice},\ }\href
  {https://doi.org/10.1103/PhysRevD.94.034504} {\bibfield  {journal} {\bibinfo
  {journal} {Phys. Rev. D}\ }\textbf {\bibinfo {volume} {94}},\ \bibinfo
  {pages} {034504} (\bibinfo {year} {2016})}\BibitemShut {NoStop}%
\bibitem [{\citenamefont {Bandyopadhyay}\ \emph {et~al.}(2020)\citenamefont
  {Bandyopadhyay}, \citenamefont {Ghosh}, \citenamefont {Farias}, \citenamefont
  {Dey},\ and\ \citenamefont {Krein}}]{PhysRevD.102.114015}%
  \BibitemOpen
  \bibfield  {author} {\bibinfo {author} {\bibfnamefont {A.}~\bibnamefont
  {Bandyopadhyay}}, \bibinfo {author} {\bibfnamefont {S.}~\bibnamefont
  {Ghosh}}, \bibinfo {author} {\bibfnamefont {R.~L.~S.}\ \bibnamefont
  {Farias}}, \bibinfo {author} {\bibfnamefont {J.}~\bibnamefont {Dey}},\ and\
  \bibinfo {author} {\bibfnamefont {G.~a.}\ \bibnamefont {Krein}},\ }\bibfield
  {title} {\bibinfo {title} {Anisotropic electrical conductivity of magnetized
  hot quark matter},\ }\href {https://doi.org/10.1103/PhysRevD.102.114015}
  {\bibfield  {journal} {\bibinfo  {journal} {Phys. Rev. D}\ }\textbf {\bibinfo
  {volume} {102}},\ \bibinfo {pages} {114015} (\bibinfo {year}
  {2020})}\BibitemShut {NoStop}%
\bibitem [{\citenamefont {Islam}\ \emph {et~al.}(2021)\citenamefont {Islam},
  \citenamefont {Dey},\ and\ \citenamefont {Ghosh}}]{PhysRevC.103.034904}%
  \BibitemOpen
  \bibfield  {author} {\bibinfo {author} {\bibfnamefont {C.~A.}\ \bibnamefont
  {Islam}}, \bibinfo {author} {\bibfnamefont {J.}~\bibnamefont {Dey}},\ and\
  \bibinfo {author} {\bibfnamefont {S.}~\bibnamefont {Ghosh}},\ }\bibfield
  {title} {\bibinfo {title} {Impact of different extended components of
  mean-field models on transport coefficients of quark matter and their causal
  aspects},\ }\href {https://doi.org/10.1103/PhysRevC.103.034904} {\bibfield
  {journal} {\bibinfo  {journal} {Phys. Rev. C}\ }\textbf {\bibinfo {volume}
  {103}},\ \bibinfo {pages} {034904} (\bibinfo {year} {2021})}\BibitemShut
  {NoStop}%
\bibitem [{\citenamefont {Dwibedi}\ \emph {et~al.}(2024)\citenamefont
  {Dwibedi}, \citenamefont {Padhan}, \citenamefont {Chatterjee},\ and\
  \citenamefont {Ghosh}}]{Dwibedi:2024mff}%
  \BibitemOpen
  \bibfield  {author} {\bibinfo {author} {\bibfnamefont {A.}~\bibnamefont
  {Dwibedi}}, \bibinfo {author} {\bibfnamefont {N.}~\bibnamefont {Padhan}},
  \bibinfo {author} {\bibfnamefont {A.}~\bibnamefont {Chatterjee}},\ and\
  \bibinfo {author} {\bibfnamefont {S.}~\bibnamefont {Ghosh}},\ }\bibfield
  {title} {\bibinfo {title} {{Transport Coefficients of Relativistic Matter: A
  Detailed Formalism with a Gross Knowledge of Their Magnitude}},\ }\href
  {https://doi.org/10.3390/universe10030132} {\bibfield  {journal} {\bibinfo
  {journal} {Universe}\ }\textbf {\bibinfo {volume} {10}},\ \bibinfo {pages}
  {132} (\bibinfo {year} {2024})},\ \Eprint {https://arxiv.org/abs/2404.01421}
  {arXiv:2404.01421 [nucl-th]} \BibitemShut {NoStop}%
\bibitem [{\citenamefont {Rai}\ \emph {et~al.}(2026)\citenamefont {Rai},
  \citenamefont {Marattukalam}, \citenamefont {Murmu}, \citenamefont {Dwibedi},
  \citenamefont {Sharma},\ and\ \citenamefont {Ghosh}}]{Rai:2025lxw}%
  \BibitemOpen
  \bibfield  {author} {\bibinfo {author} {\bibfnamefont {A.}~\bibnamefont
  {Rai}}, \bibinfo {author} {\bibfnamefont {D.~R.~J.}\ \bibnamefont
  {Marattukalam}}, \bibinfo {author} {\bibfnamefont {P.}~\bibnamefont {Murmu}},
  \bibinfo {author} {\bibfnamefont {A.}~\bibnamefont {Dwibedi}}, \bibinfo
  {author} {\bibfnamefont {R.}~\bibnamefont {Sharma}},\ and\ \bibinfo {author}
  {\bibfnamefont {S.}~\bibnamefont {Ghosh}},\ }\bibfield  {title} {\bibinfo
  {title} {{Towards compressed baryonic matter densities: thermodynamics and
  transport coefficients}},\ }\href
  {https://doi.org/10.1140/epjc/s10052-026-16111-2} {\bibfield  {journal}
  {\bibinfo  {journal} {Eur. Phys. J. C}\ }\textbf {\bibinfo {volume} {86}},\
  \bibinfo {pages} {837} (\bibinfo {year} {2026})},\ \Eprint
  {https://arxiv.org/abs/2512.20282} {arXiv:2512.20282 [nucl-th]} \BibitemShut
  {NoStop}%
\bibitem [{\citenamefont {Dey}\ \emph {et~al.}(2021)\citenamefont {Dey},
  \citenamefont {Satapathy}, \citenamefont {Mishra}, \citenamefont {Paul},\
  and\ \citenamefont {Ghosh}}]{IJMPE}%
  \BibitemOpen
  \bibfield  {author} {\bibinfo {author} {\bibfnamefont {J.}~\bibnamefont
  {Dey}}, \bibinfo {author} {\bibfnamefont {S.}~\bibnamefont {Satapathy}},
  \bibinfo {author} {\bibfnamefont {A.}~\bibnamefont {Mishra}}, \bibinfo
  {author} {\bibfnamefont {S.}~\bibnamefont {Paul}},\ and\ \bibinfo {author}
  {\bibfnamefont {S.}~\bibnamefont {Ghosh}},\ }\bibfield  {title} {\bibinfo
  {title} {From noninteracting to interacting picture of quark–gluon plasma
  in the presence of a magnetic field and its fluid property},\ }\href
  {https://doi.org/10.1142/S0218301321500440} {\bibfield  {journal} {\bibinfo
  {journal} {International Journal of Modern Physics E}\ }\textbf {\bibinfo
  {volume} {30}},\ \bibinfo {pages} {2150044} (\bibinfo {year} {2021})},\
  \Eprint {https://arxiv.org/abs/https://doi.org/10.1142/S0218301321500440}
  {https://doi.org/10.1142/S0218301321500440} \BibitemShut {NoStop}%
\bibitem [{\citenamefont {Singh}\ \emph
  {et~al.}(2024{\natexlab{a}})\citenamefont {Singh}, \citenamefont {Dey},\ and\
  \citenamefont {Sahoo}}]{PhysRevD.109.014018}%
  \BibitemOpen
  \bibfield  {author} {\bibinfo {author} {\bibfnamefont {K.}~\bibnamefont
  {Singh}}, \bibinfo {author} {\bibfnamefont {J.}~\bibnamefont {Dey}},\ and\
  \bibinfo {author} {\bibfnamefont {R.}~\bibnamefont {Sahoo}},\ }\bibfield
  {title} {\bibinfo {title} {Thermal conductivity of evolving quark-gluon
  plasma in the presence of a time-varying magnetic field},\ }\href
  {https://doi.org/10.1103/PhysRevD.109.014018} {\bibfield  {journal} {\bibinfo
   {journal} {Phys. Rev. D}\ }\textbf {\bibinfo {volume} {109}},\ \bibinfo
  {pages} {014018} (\bibinfo {year} {2024}{\natexlab{a}})}\BibitemShut
  {NoStop}%
\bibitem [{\citenamefont {Gurzhi}(1963)}]{gurzhi1963minimum}%
  \BibitemOpen
  \bibfield  {author} {\bibinfo {author} {\bibfnamefont {R.}~\bibnamefont
  {Gurzhi}},\ }\bibfield  {title} {\bibinfo {title} {Minimum of resistance in
  impurity-free conductors},\ }\href@noop {} {\bibfield  {journal} {\bibinfo
  {journal} {Sov. Phys. JETP}\ }\textbf {\bibinfo {volume} {44}},\ \bibinfo
  {pages} {771} (\bibinfo {year} {1963})}\BibitemShut {NoStop}%
\bibitem [{\citenamefont {Lucas}\ and\ \citenamefont
  {Fong}(2018)}]{Lucasfong2018}%
  \BibitemOpen
  \bibfield  {author} {\bibinfo {author} {\bibfnamefont {A.}~\bibnamefont
  {Lucas}}\ and\ \bibinfo {author} {\bibfnamefont {K.~C.}\ \bibnamefont
  {Fong}},\ }\bibfield  {title} {\bibinfo {title} {Hydrodynamics of electrons
  in graphene},\ }\href {https://doi.org/10.1088/1361-648X/aaa274} {\bibfield
  {journal} {\bibinfo  {journal} {Journal of Physics: Condensed Matter}\
  }\textbf {\bibinfo {volume} {30}},\ \bibinfo {pages} {053001} (\bibinfo
  {year} {2018})}\BibitemShut {NoStop}%
\bibitem [{\citenamefont {Bandurin}\ \emph {et~al.}(2016)\citenamefont
  {Bandurin}, \citenamefont {Torre}, \citenamefont {Kumar}, \citenamefont
  {Shalom}, \citenamefont {Tomadin}, \citenamefont {Principi}, \citenamefont
  {Auton}, \citenamefont {Khestanova}, \citenamefont {Novoselov}, \citenamefont
  {Grigorieva}, \citenamefont {Ponomarenko}, \citenamefont {Geim},\ and\
  \citenamefont {Polini}}]{Bandurin2016}%
  \BibitemOpen
  \bibfield  {author} {\bibinfo {author} {\bibfnamefont {D.~A.}\ \bibnamefont
  {Bandurin}}, \bibinfo {author} {\bibfnamefont {I.}~\bibnamefont {Torre}},
  \bibinfo {author} {\bibfnamefont {R.~K.}\ \bibnamefont {Kumar}}, \bibinfo
  {author} {\bibfnamefont {M.~B.}\ \bibnamefont {Shalom}}, \bibinfo {author}
  {\bibfnamefont {A.}~\bibnamefont {Tomadin}}, \bibinfo {author} {\bibfnamefont
  {A.}~\bibnamefont {Principi}}, \bibinfo {author} {\bibfnamefont {G.~H.}\
  \bibnamefont {Auton}}, \bibinfo {author} {\bibfnamefont {E.}~\bibnamefont
  {Khestanova}}, \bibinfo {author} {\bibfnamefont {K.~S.}\ \bibnamefont
  {Novoselov}}, \bibinfo {author} {\bibfnamefont {I.~V.}\ \bibnamefont
  {Grigorieva}}, \bibinfo {author} {\bibfnamefont {L.~A.}\ \bibnamefont
  {Ponomarenko}}, \bibinfo {author} {\bibfnamefont {A.~K.}\ \bibnamefont
  {Geim}},\ and\ \bibinfo {author} {\bibfnamefont {M.}~\bibnamefont {Polini}},\
  }\bibfield  {title} {\bibinfo {title} {Negative local resistance caused by
  viscous electron backflow in graphene},\ }\href
  {https://doi.org/10.1126/science.aad0201} {\bibfield  {journal} {\bibinfo
  {journal} {Science}\ }\textbf {\bibinfo {volume} {351}},\ \bibinfo {pages}
  {1055} (\bibinfo {year} {2016})},\ \Eprint
  {https://arxiv.org/abs/https://www.science.org/doi/pdf/10.1126/science.aad0201}
  {https://www.science.org/doi/pdf/10.1126/science.aad0201} \BibitemShut
  {NoStop}%
\bibitem [{\citenamefont {Crossno}\ \emph {et~al.}(2016)\citenamefont
  {Crossno}, \citenamefont {Shi}, \citenamefont {Wang}, \citenamefont {Liu},
  \citenamefont {Harzheim}, \citenamefont {Lucas}, \citenamefont {Sachdev},
  \citenamefont {Kim}, \citenamefont {Taniguchi}, \citenamefont {Watanabe}
  \emph {et~al.}}]{crossno2016observation}%
  \BibitemOpen
  \bibfield  {author} {\bibinfo {author} {\bibfnamefont {J.}~\bibnamefont
  {Crossno}}, \bibinfo {author} {\bibfnamefont {J.~K.}\ \bibnamefont {Shi}},
  \bibinfo {author} {\bibfnamefont {K.}~\bibnamefont {Wang}}, \bibinfo {author}
  {\bibfnamefont {X.}~\bibnamefont {Liu}}, \bibinfo {author} {\bibfnamefont
  {A.}~\bibnamefont {Harzheim}}, \bibinfo {author} {\bibfnamefont
  {A.}~\bibnamefont {Lucas}}, \bibinfo {author} {\bibfnamefont
  {S.}~\bibnamefont {Sachdev}}, \bibinfo {author} {\bibfnamefont
  {P.}~\bibnamefont {Kim}}, \bibinfo {author} {\bibfnamefont {T.}~\bibnamefont
  {Taniguchi}}, \bibinfo {author} {\bibfnamefont {K.}~\bibnamefont {Watanabe}},
  \emph {et~al.},\ }\bibfield  {title} {\bibinfo {title} {Observation of the
  dirac fluid and the breakdown of the wiedemann-franz law in graphene},\
  }\href@noop {} {\bibfield  {journal} {\bibinfo  {journal} {Science}\ }\textbf
  {\bibinfo {volume} {351}},\ \bibinfo {pages} {1058} (\bibinfo {year}
  {2016})}\BibitemShut {NoStop}%
\bibitem [{\citenamefont {Sulpizio}\ \emph {et~al.}(2019)\citenamefont
  {Sulpizio}, \citenamefont {Ella}, \citenamefont {Rozen}, \citenamefont
  {Birkbeck}, \citenamefont {Perello}, \citenamefont {Dutta}, \citenamefont
  {Ben-Shalom}, \citenamefont {Taniguchi}, \citenamefont {Watanabe},
  \citenamefont {Holder} \emph {et~al.}}]{sulpizio2019visualizing}%
  \BibitemOpen
  \bibfield  {author} {\bibinfo {author} {\bibfnamefont {J.~A.}\ \bibnamefont
  {Sulpizio}}, \bibinfo {author} {\bibfnamefont {L.}~\bibnamefont {Ella}},
  \bibinfo {author} {\bibfnamefont {A.}~\bibnamefont {Rozen}}, \bibinfo
  {author} {\bibfnamefont {J.}~\bibnamefont {Birkbeck}}, \bibinfo {author}
  {\bibfnamefont {D.~J.}\ \bibnamefont {Perello}}, \bibinfo {author}
  {\bibfnamefont {D.}~\bibnamefont {Dutta}}, \bibinfo {author} {\bibfnamefont
  {M.}~\bibnamefont {Ben-Shalom}}, \bibinfo {author} {\bibfnamefont
  {T.}~\bibnamefont {Taniguchi}}, \bibinfo {author} {\bibfnamefont
  {K.}~\bibnamefont {Watanabe}}, \bibinfo {author} {\bibfnamefont
  {T.}~\bibnamefont {Holder}}, \emph {et~al.},\ }\bibfield  {title} {\bibinfo
  {title} {Visualizing poiseuille flow of hydrodynamic electrons},\ }\href@noop
  {} {\bibfield  {journal} {\bibinfo  {journal} {Nature}\ }\textbf {\bibinfo
  {volume} {576}},\ \bibinfo {pages} {75} (\bibinfo {year} {2019})}\BibitemShut
  {NoStop}%
\bibitem [{\citenamefont {Kumar}\ \emph {et~al.}(2022)\citenamefont {Kumar},
  \citenamefont {Birkbeck}, \citenamefont {Sulpizio}, \citenamefont {Perello},
  \citenamefont {Taniguchi}, \citenamefont {Watanabe}, \citenamefont {Reuven},
  \citenamefont {Scaffidi}, \citenamefont {Stern}, \citenamefont {Geim},\ and\
  \citenamefont {Ilani}}]{Kumar2022}%
  \BibitemOpen
  \bibfield  {author} {\bibinfo {author} {\bibfnamefont {C.}~\bibnamefont
  {Kumar}}, \bibinfo {author} {\bibfnamefont {J.}~\bibnamefont {Birkbeck}},
  \bibinfo {author} {\bibfnamefont {J.~A.}\ \bibnamefont {Sulpizio}}, \bibinfo
  {author} {\bibfnamefont {D.}~\bibnamefont {Perello}}, \bibinfo {author}
  {\bibfnamefont {T.}~\bibnamefont {Taniguchi}}, \bibinfo {author}
  {\bibfnamefont {K.}~\bibnamefont {Watanabe}}, \bibinfo {author}
  {\bibfnamefont {O.}~\bibnamefont {Reuven}}, \bibinfo {author} {\bibfnamefont
  {T.}~\bibnamefont {Scaffidi}}, \bibinfo {author} {\bibfnamefont
  {A.}~\bibnamefont {Stern}}, \bibinfo {author} {\bibfnamefont {A.~K.}\
  \bibnamefont {Geim}},\ and\ \bibinfo {author} {\bibfnamefont
  {S.}~\bibnamefont {Ilani}},\ }\bibfield  {title} {\bibinfo {title} {Imaging
  hydrodynamic electrons flowing without landauer--sharvin resistance},\ }\href
  {https://doi.org/10.1038/s41586-022-05002-7} {\bibfield  {journal} {\bibinfo
  {journal} {Nature}\ }\textbf {\bibinfo {volume} {609}},\ \bibinfo {pages}
  {276} (\bibinfo {year} {2022})}\BibitemShut {NoStop}%
\bibitem [{\citenamefont {de~Jong}\ and\ \citenamefont
  {Molenkamp}(1995)}]{PhysRevB.51.13389}%
  \BibitemOpen
  \bibfield  {author} {\bibinfo {author} {\bibfnamefont {M.~J.~M.}\
  \bibnamefont {de~Jong}}\ and\ \bibinfo {author} {\bibfnamefont {L.~W.}\
  \bibnamefont {Molenkamp}},\ }\bibfield  {title} {\bibinfo {title}
  {Hydrodynamic electron flow in high-mobility wires},\ }\href
  {https://doi.org/10.1103/PhysRevB.51.13389} {\bibfield  {journal} {\bibinfo
  {journal} {Phys. Rev. B}\ }\textbf {\bibinfo {volume} {51}},\ \bibinfo
  {pages} {13389} (\bibinfo {year} {1995})}\BibitemShut {NoStop}%
\bibitem [{\citenamefont {Moll}\ \emph {et~al.}(2016)\citenamefont {Moll},
  \citenamefont {Kushwaha}, \citenamefont {Nandi}, \citenamefont {Schmidt},\
  and\ \citenamefont {Mackenzie}}]{PdCoO2}%
  \BibitemOpen
  \bibfield  {author} {\bibinfo {author} {\bibfnamefont {P.~J.~W.}\
  \bibnamefont {Moll}}, \bibinfo {author} {\bibfnamefont {P.}~\bibnamefont
  {Kushwaha}}, \bibinfo {author} {\bibfnamefont {N.}~\bibnamefont {Nandi}},
  \bibinfo {author} {\bibfnamefont {B.}~\bibnamefont {Schmidt}},\ and\ \bibinfo
  {author} {\bibfnamefont {A.~P.}\ \bibnamefont {Mackenzie}},\ }\bibfield
  {title} {\bibinfo {title} {Evidence for hydrodynamic electron flow in
  pdcoo<sub>2</sub>},\ }\href {https://doi.org/10.1126/science.aac8385}
  {\bibfield  {journal} {\bibinfo  {journal} {Science}\ }\textbf {\bibinfo
  {volume} {351}},\ \bibinfo {pages} {1061} (\bibinfo {year} {2016})},\ \Eprint
  {https://arxiv.org/abs/https://www.science.org/doi/pdf/10.1126/science.aac8385}
  {https://www.science.org/doi/pdf/10.1126/science.aac8385} \BibitemShut
  {NoStop}%
\bibitem [{\citenamefont {Gooth}\ \emph {et~al.}(2017)\citenamefont {Gooth},
  \citenamefont {Menges}, \citenamefont {Shekhar}, \citenamefont {S{\"u}{\ss}},
  \citenamefont {Kumar}, \citenamefont {Sun}, \citenamefont {Drechsler},
  \citenamefont {Zierold}, \citenamefont {Felser},\ and\ \citenamefont
  {Gotsmann}}]{gooth2017electrical}%
  \BibitemOpen
  \bibfield  {author} {\bibinfo {author} {\bibfnamefont {J.}~\bibnamefont
  {Gooth}}, \bibinfo {author} {\bibfnamefont {F.}~\bibnamefont {Menges}},
  \bibinfo {author} {\bibfnamefont {C.}~\bibnamefont {Shekhar}}, \bibinfo
  {author} {\bibfnamefont {V.}~\bibnamefont {S{\"u}{\ss}}}, \bibinfo {author}
  {\bibfnamefont {N.}~\bibnamefont {Kumar}}, \bibinfo {author} {\bibfnamefont
  {Y.}~\bibnamefont {Sun}}, \bibinfo {author} {\bibfnamefont {U.}~\bibnamefont
  {Drechsler}}, \bibinfo {author} {\bibfnamefont {R.}~\bibnamefont {Zierold}},
  \bibinfo {author} {\bibfnamefont {C.}~\bibnamefont {Felser}},\ and\ \bibinfo
  {author} {\bibfnamefont {B.}~\bibnamefont {Gotsmann}},\ }\bibfield  {title}
  {\bibinfo {title} {Electrical and thermal transport at the planckian bound of
  dissipation in the hydrodynamic electron fluid of wp2},\ }\href@noop {}
  {\bibfield  {journal} {\bibinfo  {journal} {arXiv preprint arXiv:1706.05925}\
  } (\bibinfo {year} {2017})}\BibitemShut {NoStop}%
\bibitem [{\citenamefont {Wallace}(1947)}]{Wallace1947}%
  \BibitemOpen
  \bibfield  {author} {\bibinfo {author} {\bibfnamefont {P.~R.}\ \bibnamefont
  {Wallace}},\ }\bibfield  {title} {\bibinfo {title} {The band theory of
  graphite},\ }\href {https://doi.org/10.1103/PhysRev.71.622} {\bibfield
  {journal} {\bibinfo  {journal} {Phys. Rev.}\ }\textbf {\bibinfo {volume}
  {71}},\ \bibinfo {pages} {622} (\bibinfo {year} {1947})}\BibitemShut
  {NoStop}%
\bibitem [{\citenamefont {Castro~Neto}\ \emph {et~al.}(2009)\citenamefont
  {Castro~Neto}, \citenamefont {Guinea}, \citenamefont {Peres}, \citenamefont
  {Novoselov},\ and\ \citenamefont {Geim}}]{neto2009electronic}%
  \BibitemOpen
  \bibfield  {author} {\bibinfo {author} {\bibfnamefont {A.~H.}\ \bibnamefont
  {Castro~Neto}}, \bibinfo {author} {\bibfnamefont {F.}~\bibnamefont {Guinea}},
  \bibinfo {author} {\bibfnamefont {N.~M.~R.}\ \bibnamefont {Peres}}, \bibinfo
  {author} {\bibfnamefont {K.~S.}\ \bibnamefont {Novoselov}},\ and\ \bibinfo
  {author} {\bibfnamefont {A.~K.}\ \bibnamefont {Geim}},\ }\bibfield  {title}
  {\bibinfo {title} {The electronic properties of graphene},\ }\href
  {https://doi.org/10.1103/revmodphys.81.109} {\bibfield  {journal} {\bibinfo
  {journal} {Reviews of Modern Physics}\ }\textbf {\bibinfo {volume} {81}},\
  \bibinfo {pages} {109–162} (\bibinfo {year} {2009})}\BibitemShut {NoStop}%
\bibitem [{\citenamefont {Narozhny}(2019)}]{Narozhny2019uib}%
  \BibitemOpen
  \bibfield  {author} {\bibinfo {author} {\bibfnamefont {B.~N.}\ \bibnamefont
  {Narozhny}},\ }\bibfield  {title} {\bibinfo {title} {{Electronic
  hydrodynamics in graphene}},\ }\href
  {https://doi.org/10.1016/j.aop.2019.167979} {\bibfield  {journal} {\bibinfo
  {journal} {Annals Phys.}\ }\textbf {\bibinfo {volume} {411}},\ \bibinfo
  {pages} {167979} (\bibinfo {year} {2019})},\ \Eprint
  {https://arxiv.org/abs/1905.09686} {arXiv:1905.09686 [cond-mat.mes-hall]}
  \BibitemShut {NoStop}%
\bibitem [{\citenamefont {Win}\ \emph {et~al.}(2025)\citenamefont {Win},
  \citenamefont {Aung}, \citenamefont {Khandal},\ and\ \citenamefont
  {Ghosh}}]{win2025graphene}%
  \BibitemOpen
  \bibfield  {author} {\bibinfo {author} {\bibfnamefont {T.~Z.}\ \bibnamefont
  {Win}}, \bibinfo {author} {\bibfnamefont {C.~W.}\ \bibnamefont {Aung}},
  \bibinfo {author} {\bibfnamefont {G.}~\bibnamefont {Khandal}},\ and\ \bibinfo
  {author} {\bibfnamefont {S.}~\bibnamefont {Ghosh}},\ }\bibfield  {title}
  {\bibinfo {title} {Graphene is neither relativistic nor non-relativistic:
  thermodynamics aspects},\ }\href {https://doi.org/10.1007/s12043-024-02888-y}
  {\bibfield  {journal} {\bibinfo  {journal} {Pramana}\ }\textbf {\bibinfo
  {volume} {99}},\ \bibinfo {pages} {28} (\bibinfo {year} {2025})}\BibitemShut
  {NoStop}%
\bibitem [{\citenamefont {Narozhny}(2022)}]{Narozhny:2022ncn}%
  \BibitemOpen
  \bibfield  {author} {\bibinfo {author} {\bibfnamefont {B.~N.}\ \bibnamefont
  {Narozhny}},\ }\bibfield  {title} {\bibinfo {title} {{Hydrodynamic approach
  to two-dimensional electron systems}},\ }\href
  {https://doi.org/10.1007/s40766-022-00036-z} {\bibfield  {journal} {\bibinfo
  {journal} {Riv. Nuovo Cim.}\ }\textbf {\bibinfo {volume} {45}},\ \bibinfo
  {pages} {661} (\bibinfo {year} {2022})},\ \Eprint
  {https://arxiv.org/abs/2207.10004} {arXiv:2207.10004 [cond-mat.mes-hall]}
  \BibitemShut {NoStop}%
\bibitem [{\citenamefont {Novoselov}(2011)}]{RevModPhys.83.837}%
  \BibitemOpen
  \bibfield  {author} {\bibinfo {author} {\bibfnamefont {K.~S.}\ \bibnamefont
  {Novoselov}},\ }\bibfield  {title} {\bibinfo {title} {Nobel lecture:
  Graphene: Materials in the flatland},\ }\href
  {https://doi.org/10.1103/RevModPhys.83.837} {\bibfield  {journal} {\bibinfo
  {journal} {Rev. Mod. Phys.}\ }\textbf {\bibinfo {volume} {83}},\ \bibinfo
  {pages} {837} (\bibinfo {year} {2011})}\BibitemShut {NoStop}%
\bibitem [{\citenamefont {Geim}(2011)}]{RevModPhys.83.851}%
  \BibitemOpen
  \bibfield  {author} {\bibinfo {author} {\bibfnamefont {A.~K.}\ \bibnamefont
  {Geim}},\ }\bibfield  {title} {\bibinfo {title} {Nobel lecture: Random walk
  to graphene},\ }\href {https://doi.org/10.1103/RevModPhys.83.851} {\bibfield
  {journal} {\bibinfo  {journal} {Rev. Mod. Phys.}\ }\textbf {\bibinfo {volume}
  {83}},\ \bibinfo {pages} {851} (\bibinfo {year} {2011})}\BibitemShut
  {NoStop}%
\bibitem [{\citenamefont {Nayak}\ \emph {et~al.}(2025)\citenamefont {Nayak},
  \citenamefont {Aung}, \citenamefont {Win}, \citenamefont {Dwibedi},
  \citenamefont {Ghosh},\ and\ \citenamefont
  {Vempati}}]{nayak2025electronhydrodynamicsgraphene}%
  \BibitemOpen
  \bibfield  {author} {\bibinfo {author} {\bibfnamefont {S.}~\bibnamefont
  {Nayak}}, \bibinfo {author} {\bibfnamefont {C.~W.}\ \bibnamefont {Aung}},
  \bibinfo {author} {\bibfnamefont {T.~Z.}\ \bibnamefont {Win}}, \bibinfo
  {author} {\bibfnamefont {A.}~\bibnamefont {Dwibedi}}, \bibinfo {author}
  {\bibfnamefont {S.}~\bibnamefont {Ghosh}},\ and\ \bibinfo {author}
  {\bibfnamefont {S.}~\bibnamefont {Vempati}},\ }\href
  {https://arxiv.org/abs/2509.11315} {\bibinfo {title} {Electron hydrodynamics
  in graphene : Experimental and theoretical status}} (\bibinfo {year}
  {2025}),\ \Eprint {https://arxiv.org/abs/2509.11315} {arXiv:2509.11315
  [cond-mat.mes-hall]} \BibitemShut {NoStop}%
\bibitem [{\citenamefont {Fritz}\ and\ \citenamefont
  {Scaffidi}(2024)}]{Fritz2023Hydrodynamic}%
  \BibitemOpen
  \bibfield  {author} {\bibinfo {author} {\bibfnamefont {L.}~\bibnamefont
  {Fritz}}\ and\ \bibinfo {author} {\bibfnamefont {T.}~\bibnamefont
  {Scaffidi}},\ }\bibfield  {title} {\bibinfo {title} {Hydrodynamic electronic
  transport},\ }\href
  {https://doi.org/10.1146/annurev-conmatphys-040521-042014} {\bibfield
  {journal} {\bibinfo  {journal} {Annual Review of Condensed Matter Physics}\
  }\textbf {\bibinfo {volume} {15}},\ \bibinfo {pages} {139} (\bibinfo {year}
  {2024})},\ \Eprint {https://arxiv.org/abs/2303.14205} {arXiv:2303.14205
  [cond-mat.str-el]} \BibitemShut {NoStop}%
\bibitem [{\citenamefont {Hui}\ and\ \citenamefont {Skinner}(2025)}]{Hui_2025}%
  \BibitemOpen
  \bibfield  {author} {\bibinfo {author} {\bibfnamefont {A.}~\bibnamefont
  {Hui}}\ and\ \bibinfo {author} {\bibfnamefont {B.}~\bibnamefont {Skinner}},\
  }\bibfield  {title} {\bibinfo {title} {Hydrodynamics of the electronic fermi
  liquid: a pedagogical overview},\ }\href
  {https://doi.org/10.1088/1361-648X/adfbcd} {\bibfield  {journal} {\bibinfo
  {journal} {Journal of Physics: Condensed Matter}\ }\textbf {\bibinfo {volume}
  {37}},\ \bibinfo {pages} {363001} (\bibinfo {year} {2025})}\BibitemShut
  {NoStop}%
\bibitem [{\citenamefont {Dwibedi}\ \emph {et~al.}(2025)\citenamefont
  {Dwibedi}, \citenamefont {Nayak}, \citenamefont {Kiran}, \citenamefont
  {Ghosh},\ and\ \citenamefont {Vempati}}]{Dwibedi2025}%
  \BibitemOpen
  \bibfield  {author} {\bibinfo {author} {\bibfnamefont {A.}~\bibnamefont
  {Dwibedi}}, \bibinfo {author} {\bibfnamefont {S.}~\bibnamefont {Nayak}},
  \bibinfo {author} {\bibfnamefont {S.~S.}\ \bibnamefont {Kiran}}, \bibinfo
  {author} {\bibfnamefont {S.}~\bibnamefont {Ghosh}},\ and\ \bibinfo {author}
  {\bibfnamefont {S.}~\bibnamefont {Vempati}},\ }\bibfield  {title} {\bibinfo
  {title} {On the wiedemann--franz law violation in graphene and quark--gluon
  plasma systems},\ }\href {https://doi.org/10.1140/epjb/s10051-025-01009-x}
  {\bibfield  {journal} {\bibinfo  {journal} {The European Physical Journal B}\
  }\textbf {\bibinfo {volume} {98}},\ \bibinfo {pages} {167} (\bibinfo {year}
  {2025})}\BibitemShut {NoStop}%
\bibitem [{\citenamefont {Win}\ \emph {et~al.}(0)\citenamefont {Win},
  \citenamefont {Aung}, \citenamefont {Khandal},\ and\ \citenamefont
  {Ghosh}}]{win2024wied}%
  \BibitemOpen
  \bibfield  {author} {\bibinfo {author} {\bibfnamefont {T.~Z.}\ \bibnamefont
  {Win}}, \bibinfo {author} {\bibfnamefont {C.~W.}\ \bibnamefont {Aung}},
  \bibinfo {author} {\bibfnamefont {G.}~\bibnamefont {Khandal}},\ and\ \bibinfo
  {author} {\bibfnamefont {S.}~\bibnamefont {Ghosh}},\ }\bibfield  {title}
  {\bibinfo {title} {Wiedemann–franz law violation domain for graphene and
  nonrelativistic systems},\ }\href {https://doi.org/10.1142/S0217979225501826}
  {\bibfield  {journal} {\bibinfo  {journal} {International Journal of Modern
  Physics B}\ }\textbf {\bibinfo {volume} {0}},\ \bibinfo {pages} {2550182}
  (\bibinfo {year} {0})},\ \Eprint
  {https://arxiv.org/abs/https://doi.org/10.1142/S0217979225501826}
  {https://doi.org/10.1142/S0217979225501826} \BibitemShut {NoStop}%
\bibitem [{\citenamefont {Majumdar}\ \emph {et~al.}(2025)\citenamefont
  {Majumdar}, \citenamefont {Chadha}, \citenamefont {Pal}, \citenamefont
  {Gugnani}, \citenamefont {Ghawri}, \citenamefont {Watanabe}, \citenamefont
  {Taniguchi}, \citenamefont {Mukerjee},\ and\ \citenamefont
  {Ghosh}}]{majumdar2025universality}%
  \BibitemOpen
  \bibfield  {author} {\bibinfo {author} {\bibfnamefont {A.}~\bibnamefont
  {Majumdar}}, \bibinfo {author} {\bibfnamefont {N.}~\bibnamefont {Chadha}},
  \bibinfo {author} {\bibfnamefont {P.}~\bibnamefont {Pal}}, \bibinfo {author}
  {\bibfnamefont {A.}~\bibnamefont {Gugnani}}, \bibinfo {author} {\bibfnamefont
  {B.}~\bibnamefont {Ghawri}}, \bibinfo {author} {\bibfnamefont
  {K.}~\bibnamefont {Watanabe}}, \bibinfo {author} {\bibfnamefont
  {T.}~\bibnamefont {Taniguchi}}, \bibinfo {author} {\bibfnamefont
  {S.}~\bibnamefont {Mukerjee}},\ and\ \bibinfo {author} {\bibfnamefont
  {A.}~\bibnamefont {Ghosh}},\ }\href {https://arxiv.org/abs/2501.03193}
  {\bibinfo {title} {Universality in quantum critical flow of charge and heat
  in ultra-clean graphene}} (\bibinfo {year} {2025}),\ \Eprint
  {https://arxiv.org/abs/2501.03193} {arXiv:2501.03193 [cond-mat.mes-hall]}
  \BibitemShut {NoStop}%
\bibitem [{\citenamefont {Rath}\ \emph {et~al.}(2019)\citenamefont {Rath},
  \citenamefont {Tripathy}, \citenamefont {Chatterjee}, \citenamefont {Sahoo},
  \citenamefont {Kumar~Tiwari},\ and\ \citenamefont {Nath}}]{Rath:2019nne}%
  \BibitemOpen
  \bibfield  {author} {\bibinfo {author} {\bibfnamefont {R.}~\bibnamefont
  {Rath}}, \bibinfo {author} {\bibfnamefont {S.}~\bibnamefont {Tripathy}},
  \bibinfo {author} {\bibfnamefont {B.}~\bibnamefont {Chatterjee}}, \bibinfo
  {author} {\bibfnamefont {R.}~\bibnamefont {Sahoo}}, \bibinfo {author}
  {\bibfnamefont {S.}~\bibnamefont {Kumar~Tiwari}},\ and\ \bibinfo {author}
  {\bibfnamefont {A.}~\bibnamefont {Nath}},\ }\bibfield  {title} {\bibinfo
  {title} {{Violation of Wiedemann-Franz Law for Hot Hadronic Matter created at
  NICA, FAIR and RHIC Energies using Non-extensive Statistics}},\ }\href
  {https://doi.org/10.1140/epja/i2019-12814-3} {\bibfield  {journal} {\bibinfo
  {journal} {Eur. Phys. J. A}\ }\textbf {\bibinfo {volume} {55}},\ \bibinfo
  {pages} {125} (\bibinfo {year} {2019})},\ \Eprint
  {https://arxiv.org/abs/1902.07922} {arXiv:1902.07922 [hep-ph]} \BibitemShut
  {NoStop}%
\bibitem [{\citenamefont {Sahoo}\ \emph {et~al.}(2019)\citenamefont {Sahoo},
  \citenamefont {Sahoo},\ and\ \citenamefont {Tiwari}}]{Sahoo:2019xjq}%
  \BibitemOpen
  \bibfield  {author} {\bibinfo {author} {\bibfnamefont {P.}~\bibnamefont
  {Sahoo}}, \bibinfo {author} {\bibfnamefont {R.}~\bibnamefont {Sahoo}},\ and\
  \bibinfo {author} {\bibfnamefont {S.~K.}\ \bibnamefont {Tiwari}},\ }\bibfield
   {title} {\bibinfo {title} {{Wiedemann-Franz law for hot QCD matter in a
  color string percolation scenario}},\ }\href
  {https://doi.org/10.1103/PhysRevD.100.051503} {\bibfield  {journal} {\bibinfo
   {journal} {Phys. Rev. D}\ }\textbf {\bibinfo {volume} {100}},\ \bibinfo
  {pages} {051503} (\bibinfo {year} {2019})},\ \Eprint
  {https://arxiv.org/abs/1904.06961} {arXiv:1904.06961 [hep-ph]} \BibitemShut
  {NoStop}%
\bibitem [{\citenamefont {Singh}\ \emph {et~al.}(2023)\citenamefont {Singh},
  \citenamefont {Dey}, \citenamefont {Sahoo},\ and\ \citenamefont
  {Ghosh}}]{PhysRevD.108.094007}%
  \BibitemOpen
  \bibfield  {author} {\bibinfo {author} {\bibfnamefont {K.}~\bibnamefont
  {Singh}}, \bibinfo {author} {\bibfnamefont {J.}~\bibnamefont {Dey}}, \bibinfo
  {author} {\bibfnamefont {R.}~\bibnamefont {Sahoo}},\ and\ \bibinfo {author}
  {\bibfnamefont {S.}~\bibnamefont {Ghosh}},\ }\bibfield  {title} {\bibinfo
  {title} {Effect of time-varying electromagnetic field on wiedemann-franz law
  in a hot hadronic matter},\ }\href
  {https://doi.org/10.1103/PhysRevD.108.094007} {\bibfield  {journal} {\bibinfo
   {journal} {Phys. Rev. D}\ }\textbf {\bibinfo {volume} {108}},\ \bibinfo
  {pages} {094007} (\bibinfo {year} {2023})}\BibitemShut {NoStop}%
\bibitem [{\citenamefont {Pradhan}\ \emph {et~al.}(2023)\citenamefont
  {Pradhan}, \citenamefont {Sahu}, \citenamefont {Scaria},\ and\ \citenamefont
  {Sahoo}}]{pradhan2023conductivity}%
  \BibitemOpen
  \bibfield  {author} {\bibinfo {author} {\bibfnamefont {K.~K.}\ \bibnamefont
  {Pradhan}}, \bibinfo {author} {\bibfnamefont {D.}~\bibnamefont {Sahu}},
  \bibinfo {author} {\bibfnamefont {R.}~\bibnamefont {Scaria}},\ and\ \bibinfo
  {author} {\bibfnamefont {R.}~\bibnamefont {Sahoo}},\ }\bibfield  {title}
  {\bibinfo {title} {Conductivity, diffusivity, and violation of the
  wiedemann-franz law in a hadron resonance gas with van der waals
  interactions},\ }\href {https://doi.org/10.1103/PhysRevC.107.014910}
  {\bibfield  {journal} {\bibinfo  {journal} {Phys. Rev. C}\ }\textbf {\bibinfo
  {volume} {107}},\ \bibinfo {pages} {014910} (\bibinfo {year}
  {2023})}\BibitemShut {NoStop}%
\bibitem [{\citenamefont {Ashcroft}\ and\ \citenamefont
  {Mermin}(1993)}]{ashcroft1993solid}%
  \BibitemOpen
  \bibfield  {author} {\bibinfo {author} {\bibfnamefont {N.}~\bibnamefont
  {Ashcroft}}\ and\ \bibinfo {author} {\bibfnamefont {N.}~\bibnamefont
  {Mermin}},\ }\bibfield  {title} {\bibinfo {title} {Solid state physics
  (brooks cole, 1976)},\ }\href@noop {} {\bibfield  {journal} {\bibinfo
  {journal} {Cited on}\ }\textbf {\bibinfo {volume} {26}},\ \bibinfo {pages}
  {49} (\bibinfo {year} {1993})}\BibitemShut {NoStop}%
\bibitem [{\citenamefont {Zuev}\ \emph {et~al.}(2009)\citenamefont {Zuev},
  \citenamefont {Chang},\ and\ \citenamefont {Kim}}]{PhysRevLett.102.096807}%
  \BibitemOpen
  \bibfield  {author} {\bibinfo {author} {\bibfnamefont {Y.~M.}\ \bibnamefont
  {Zuev}}, \bibinfo {author} {\bibfnamefont {W.}~\bibnamefont {Chang}},\ and\
  \bibinfo {author} {\bibfnamefont {P.}~\bibnamefont {Kim}},\ }\bibfield
  {title} {\bibinfo {title} {Thermoelectric and magnetothermoelectric transport
  measurements of graphene},\ }\href
  {https://doi.org/10.1103/PhysRevLett.102.096807} {\bibfield  {journal}
  {\bibinfo  {journal} {Phys. Rev. Lett.}\ }\textbf {\bibinfo {volume} {102}},\
  \bibinfo {pages} {096807} (\bibinfo {year} {2009})}\BibitemShut {NoStop}%
\bibitem [{\citenamefont {Ghahari}\ \emph {et~al.}(2016)\citenamefont
  {Ghahari}, \citenamefont {Xie}, \citenamefont {Taniguchi}, \citenamefont
  {Watanabe}, \citenamefont {Foster},\ and\ \citenamefont
  {Kim}}]{PhysRevLett.116.136802}%
  \BibitemOpen
  \bibfield  {author} {\bibinfo {author} {\bibfnamefont {F.}~\bibnamefont
  {Ghahari}}, \bibinfo {author} {\bibfnamefont {H.-Y.}\ \bibnamefont {Xie}},
  \bibinfo {author} {\bibfnamefont {T.}~\bibnamefont {Taniguchi}}, \bibinfo
  {author} {\bibfnamefont {K.}~\bibnamefont {Watanabe}}, \bibinfo {author}
  {\bibfnamefont {M.~S.}\ \bibnamefont {Foster}},\ and\ \bibinfo {author}
  {\bibfnamefont {P.}~\bibnamefont {Kim}},\ }\bibfield  {title} {\bibinfo
  {title} {Enhanced thermoelectric power in graphene: Violation of the mott
  relation by inelastic scattering},\ }\href
  {https://doi.org/10.1103/PhysRevLett.116.136802} {\bibfield  {journal}
  {\bibinfo  {journal} {Phys. Rev. Lett.}\ }\textbf {\bibinfo {volume} {116}},\
  \bibinfo {pages} {136802} (\bibinfo {year} {2016})}\BibitemShut {NoStop}%
\bibitem [{\citenamefont {Guarochico-Moreira}\ \emph
  {et~al.}(2023)\citenamefont {Guarochico-Moreira}, \citenamefont {Anderson},
  \citenamefont {Fal'ko}, \citenamefont {Grigorieva}, \citenamefont
  {T\'ov\'ari}, \citenamefont {Hamer}, \citenamefont {Gorbachev}, \citenamefont
  {Liu}, \citenamefont {Edgar}, \citenamefont {Principi}, \citenamefont
  {Kretinin},\ and\ \citenamefont {Vera-Marun}}]{PhysRevB.108.115418}%
  \BibitemOpen
  \bibfield  {author} {\bibinfo {author} {\bibfnamefont {V.~H.}\ \bibnamefont
  {Guarochico-Moreira}}, \bibinfo {author} {\bibfnamefont {C.~R.}\ \bibnamefont
  {Anderson}}, \bibinfo {author} {\bibfnamefont {V.}~\bibnamefont {Fal'ko}},
  \bibinfo {author} {\bibfnamefont {I.~V.}\ \bibnamefont {Grigorieva}},
  \bibinfo {author} {\bibfnamefont {E.}~\bibnamefont {T\'ov\'ari}}, \bibinfo
  {author} {\bibfnamefont {M.}~\bibnamefont {Hamer}}, \bibinfo {author}
  {\bibfnamefont {R.}~\bibnamefont {Gorbachev}}, \bibinfo {author}
  {\bibfnamefont {S.}~\bibnamefont {Liu}}, \bibinfo {author} {\bibfnamefont
  {J.~H.}\ \bibnamefont {Edgar}}, \bibinfo {author} {\bibfnamefont
  {A.}~\bibnamefont {Principi}}, \bibinfo {author} {\bibfnamefont {A.~V.}\
  \bibnamefont {Kretinin}},\ and\ \bibinfo {author} {\bibfnamefont {I.~J.}\
  \bibnamefont {Vera-Marun}},\ }\bibfield  {title} {\bibinfo {title}
  {Thermopower in hbn/graphene/hbn superlattices},\ }\href
  {https://doi.org/10.1103/PhysRevB.108.115418} {\bibfield  {journal} {\bibinfo
   {journal} {Phys. Rev. B}\ }\textbf {\bibinfo {volume} {108}},\ \bibinfo
  {pages} {115418} (\bibinfo {year} {2023})}\BibitemShut {NoStop}%
\bibitem [{\citenamefont {Wei}\ \emph {et~al.}(2009)\citenamefont {Wei},
  \citenamefont {Bao}, \citenamefont {Pu}, \citenamefont {Lau},\ and\
  \citenamefont {Shi}}]{PhysRevLett.102.166808}%
  \BibitemOpen
  \bibfield  {author} {\bibinfo {author} {\bibfnamefont {P.}~\bibnamefont
  {Wei}}, \bibinfo {author} {\bibfnamefont {W.}~\bibnamefont {Bao}}, \bibinfo
  {author} {\bibfnamefont {Y.}~\bibnamefont {Pu}}, \bibinfo {author}
  {\bibfnamefont {C.~N.}\ \bibnamefont {Lau}},\ and\ \bibinfo {author}
  {\bibfnamefont {J.}~\bibnamefont {Shi}},\ }\bibfield  {title} {\bibinfo
  {title} {Anomalous thermoelectric transport of dirac particles in graphene},\
  }\href {https://doi.org/10.1103/PhysRevLett.102.166808} {\bibfield  {journal}
  {\bibinfo  {journal} {Phys. Rev. Lett.}\ }\textbf {\bibinfo {volume} {102}},\
  \bibinfo {pages} {166808} (\bibinfo {year} {2009})}\BibitemShut {NoStop}%
\bibitem [{\citenamefont {Cutler}\ and\ \citenamefont
  {Mott}(1969)}]{PhysRev.181.1336}%
  \BibitemOpen
  \bibfield  {author} {\bibinfo {author} {\bibfnamefont {M.}~\bibnamefont
  {Cutler}}\ and\ \bibinfo {author} {\bibfnamefont {N.~F.}\ \bibnamefont
  {Mott}},\ }\bibfield  {title} {\bibinfo {title} {Observation of anderson
  localization in an electron gas},\ }\href
  {https://doi.org/10.1103/PhysRev.181.1336} {\bibfield  {journal} {\bibinfo
  {journal} {Phys. Rev.}\ }\textbf {\bibinfo {volume} {181}},\ \bibinfo {pages}
  {1336} (\bibinfo {year} {1969})}\BibitemShut {NoStop}%
\bibitem [{\citenamefont {Mahapatra}\ \emph {et~al.}(2020)\citenamefont
  {Mahapatra}, \citenamefont {Ghawri}, \citenamefont {Garg}, \citenamefont
  {Mandal}, \citenamefont {Watanabe}, \citenamefont {Taniguchi}, \citenamefont
  {Jain}, \citenamefont {Mukerjee},\ and\ \citenamefont
  {Ghosh}}]{PhysRevLett.125.226802}%
  \BibitemOpen
  \bibfield  {author} {\bibinfo {author} {\bibfnamefont {P.~S.}\ \bibnamefont
  {Mahapatra}}, \bibinfo {author} {\bibfnamefont {B.}~\bibnamefont {Ghawri}},
  \bibinfo {author} {\bibfnamefont {M.}~\bibnamefont {Garg}}, \bibinfo {author}
  {\bibfnamefont {S.}~\bibnamefont {Mandal}}, \bibinfo {author} {\bibfnamefont
  {K.}~\bibnamefont {Watanabe}}, \bibinfo {author} {\bibfnamefont
  {T.}~\bibnamefont {Taniguchi}}, \bibinfo {author} {\bibfnamefont
  {M.}~\bibnamefont {Jain}}, \bibinfo {author} {\bibfnamefont {S.}~\bibnamefont
  {Mukerjee}},\ and\ \bibinfo {author} {\bibfnamefont {A.}~\bibnamefont
  {Ghosh}},\ }\bibfield  {title} {\bibinfo {title} {Misorientation-controlled
  cross-plane thermoelectricity in twisted bilayer graphene},\ }\href
  {https://doi.org/10.1103/PhysRevLett.125.226802} {\bibfield  {journal}
  {\bibinfo  {journal} {Phys. Rev. Lett.}\ }\textbf {\bibinfo {volume} {125}},\
  \bibinfo {pages} {226802} (\bibinfo {year} {2020})}\BibitemShut {NoStop}%
\bibitem [{\citenamefont {Paul}\ \emph {et~al.}(2022)\citenamefont {Paul},
  \citenamefont {Ghosh}, \citenamefont {Chakraborty}, \citenamefont {Roy},
  \citenamefont {Dutta}, \citenamefont {Watanabe}, \citenamefont {Taniguchi},
  \citenamefont {Panda}, \citenamefont {Agarwala}, \citenamefont {Mukerjee},
  \citenamefont {Banerjee},\ and\ \citenamefont {Das}}]{Paul2022}%
  \BibitemOpen
  \bibfield  {author} {\bibinfo {author} {\bibfnamefont {A.~K.}\ \bibnamefont
  {Paul}}, \bibinfo {author} {\bibfnamefont {A.}~\bibnamefont {Ghosh}},
  \bibinfo {author} {\bibfnamefont {S.}~\bibnamefont {Chakraborty}}, \bibinfo
  {author} {\bibfnamefont {U.}~\bibnamefont {Roy}}, \bibinfo {author}
  {\bibfnamefont {R.}~\bibnamefont {Dutta}}, \bibinfo {author} {\bibfnamefont
  {K.}~\bibnamefont {Watanabe}}, \bibinfo {author} {\bibfnamefont
  {T.}~\bibnamefont {Taniguchi}}, \bibinfo {author} {\bibfnamefont
  {A.}~\bibnamefont {Panda}}, \bibinfo {author} {\bibfnamefont
  {A.}~\bibnamefont {Agarwala}}, \bibinfo {author} {\bibfnamefont
  {S.}~\bibnamefont {Mukerjee}}, \bibinfo {author} {\bibfnamefont
  {S.}~\bibnamefont {Banerjee}},\ and\ \bibinfo {author} {\bibfnamefont
  {A.}~\bibnamefont {Das}},\ }\bibfield  {title} {\bibinfo {title}
  {Interaction-driven giant thermopower in magic-angle twisted bilayer
  graphene},\ }\href {https://doi.org/10.1038/s41567-022-01574-3} {\bibfield
  {journal} {\bibinfo  {journal} {Nature Physics}\ }\textbf {\bibinfo {volume}
  {18}},\ \bibinfo {pages} {691} (\bibinfo {year} {2022})}\BibitemShut
  {NoStop}%
\bibitem [{\citenamefont {Ghawri}\ \emph {et~al.}(2022)\citenamefont {Ghawri},
  \citenamefont {Mahapatra}, \citenamefont {Garg}, \citenamefont {Mandal},
  \citenamefont {Bhowmik}, \citenamefont {Jayaraman}, \citenamefont {Soni},
  \citenamefont {Watanabe}, \citenamefont {Taniguchi}, \citenamefont
  {Krishnamurthy} \emph {et~al.}}]{ghawri2022breakdown}%
  \BibitemOpen
  \bibfield  {author} {\bibinfo {author} {\bibfnamefont {B.}~\bibnamefont
  {Ghawri}}, \bibinfo {author} {\bibfnamefont {P.~S.}\ \bibnamefont
  {Mahapatra}}, \bibinfo {author} {\bibfnamefont {M.}~\bibnamefont {Garg}},
  \bibinfo {author} {\bibfnamefont {S.}~\bibnamefont {Mandal}}, \bibinfo
  {author} {\bibfnamefont {S.}~\bibnamefont {Bhowmik}}, \bibinfo {author}
  {\bibfnamefont {A.}~\bibnamefont {Jayaraman}}, \bibinfo {author}
  {\bibfnamefont {R.}~\bibnamefont {Soni}}, \bibinfo {author} {\bibfnamefont
  {K.}~\bibnamefont {Watanabe}}, \bibinfo {author} {\bibfnamefont
  {T.}~\bibnamefont {Taniguchi}}, \bibinfo {author} {\bibfnamefont
  {H.}~\bibnamefont {Krishnamurthy}}, \emph {et~al.},\ }\bibfield  {title}
  {\bibinfo {title} {Breakdown of semiclassical description of
  thermoelectricity in near-magic angle twisted bilayer graphene},\ }\href
  {https://doi.org/10.1038/s41467-022-29198-4} {\bibfield  {journal} {\bibinfo
  {journal} {Nature Communications}\ }\textbf {\bibinfo {volume} {13}},\
  \bibinfo {pages} {1522} (\bibinfo {year} {2022})}\BibitemShut {NoStop}%
\bibitem [{\citenamefont {Duan}\ \emph {et~al.}(2016)\citenamefont {Duan},
  \citenamefont {Wang}, \citenamefont {Lai}, \citenamefont {Li}, \citenamefont
  {Watanabe}, \citenamefont {Taniguchi}, \citenamefont {Zebarjadi},\ and\
  \citenamefont {Andrei}}]{doi:10.1073/pnas.1615913113}%
  \BibitemOpen
  \bibfield  {author} {\bibinfo {author} {\bibfnamefont {J.}~\bibnamefont
  {Duan}}, \bibinfo {author} {\bibfnamefont {X.}~\bibnamefont {Wang}}, \bibinfo
  {author} {\bibfnamefont {X.}~\bibnamefont {Lai}}, \bibinfo {author}
  {\bibfnamefont {G.}~\bibnamefont {Li}}, \bibinfo {author} {\bibfnamefont
  {K.}~\bibnamefont {Watanabe}}, \bibinfo {author} {\bibfnamefont
  {T.}~\bibnamefont {Taniguchi}}, \bibinfo {author} {\bibfnamefont
  {M.}~\bibnamefont {Zebarjadi}},\ and\ \bibinfo {author} {\bibfnamefont
  {E.~Y.}\ \bibnamefont {Andrei}},\ }\bibfield  {title} {\bibinfo {title} {High
  thermoelectricpower factor in graphene/hbn devices},\ }\href
  {https://doi.org/10.1073/pnas.1615913113} {\bibfield  {journal} {\bibinfo
  {journal} {Proceedings of the National Academy of Sciences}\ }\textbf
  {\bibinfo {volume} {113}},\ \bibinfo {pages} {14272} (\bibinfo {year}
  {2016})},\ \Eprint
  {https://arxiv.org/abs/https://www.pnas.org/doi/pdf/10.1073/pnas.1615913113}
  {https://www.pnas.org/doi/pdf/10.1073/pnas.1615913113} \BibitemShut {NoStop}%
\bibitem [{\citenamefont {L\"ofwander}\ and\ \citenamefont
  {Fogelstr\"om}(2007)}]{PhysRevB.76.193401}%
  \BibitemOpen
  \bibfield  {author} {\bibinfo {author} {\bibfnamefont {T.}~\bibnamefont
  {L\"ofwander}}\ and\ \bibinfo {author} {\bibfnamefont {M.}~\bibnamefont
  {Fogelstr\"om}},\ }\bibfield  {title} {\bibinfo {title} {Impurity scattering
  and mott's formula in graphene},\ }\href
  {https://doi.org/10.1103/PhysRevB.76.193401} {\bibfield  {journal} {\bibinfo
  {journal} {Phys. Rev. B}\ }\textbf {\bibinfo {volume} {76}},\ \bibinfo
  {pages} {193401} (\bibinfo {year} {2007})}\BibitemShut {NoStop}%
\bibitem [{\citenamefont {Foster}\ and\ \citenamefont
  {Aleiner}(2009)}]{PhysRevB.79.085415}%
  \BibitemOpen
  \bibfield  {author} {\bibinfo {author} {\bibfnamefont {M.~S.}\ \bibnamefont
  {Foster}}\ and\ \bibinfo {author} {\bibfnamefont {I.~L.}\ \bibnamefont
  {Aleiner}},\ }\bibfield  {title} {\bibinfo {title} {Slow imbalance relaxation
  and thermoelectric transport in graphene},\ }\href
  {https://doi.org/10.1103/PhysRevB.79.085415} {\bibfield  {journal} {\bibinfo
  {journal} {Phys. Rev. B}\ }\textbf {\bibinfo {volume} {79}},\ \bibinfo
  {pages} {085415} (\bibinfo {year} {2009})}\BibitemShut {NoStop}%
\bibitem [{\citenamefont {Hwang}\ \emph {et~al.}(2009)\citenamefont {Hwang},
  \citenamefont {Rossi},\ and\ \citenamefont {Das~Sarma}}]{PhysRevB.80.235415}%
  \BibitemOpen
  \bibfield  {author} {\bibinfo {author} {\bibfnamefont {E.~H.}\ \bibnamefont
  {Hwang}}, \bibinfo {author} {\bibfnamefont {E.}~\bibnamefont {Rossi}},\ and\
  \bibinfo {author} {\bibfnamefont {S.}~\bibnamefont {Das~Sarma}},\ }\bibfield
  {title} {\bibinfo {title} {Theory of thermopower in two-dimensional
  graphene},\ }\href {https://doi.org/10.1103/PhysRevB.80.235415} {\bibfield
  {journal} {\bibinfo  {journal} {Phys. Rev. B}\ }\textbf {\bibinfo {volume}
  {80}},\ \bibinfo {pages} {235415} (\bibinfo {year} {2009})}\BibitemShut
  {NoStop}%
\bibitem [{\citenamefont {Xie}\ and\ \citenamefont
  {Foster}(2016)}]{PhysRevB.93.195103}%
  \BibitemOpen
  \bibfield  {author} {\bibinfo {author} {\bibfnamefont {H.-Y.}\ \bibnamefont
  {Xie}}\ and\ \bibinfo {author} {\bibfnamefont {M.~S.}\ \bibnamefont
  {Foster}},\ }\bibfield  {title} {\bibinfo {title} {Transport coefficients of
  graphene: Interplay of impurity scattering, coulomb interaction, and optical
  phonons},\ }\href {https://doi.org/10.1103/PhysRevB.93.195103} {\bibfield
  {journal} {\bibinfo  {journal} {Phys. Rev. B}\ }\textbf {\bibinfo {volume}
  {93}},\ \bibinfo {pages} {195103} (\bibinfo {year} {2016})}\BibitemShut
  {NoStop}%
\bibitem [{\citenamefont {Lucas}\ and\ \citenamefont
  {Das~Sarma}(2018)}]{PhysRevB.97.245128}%
  \BibitemOpen
  \bibfield  {author} {\bibinfo {author} {\bibfnamefont {A.}~\bibnamefont
  {Lucas}}\ and\ \bibinfo {author} {\bibfnamefont {S.}~\bibnamefont
  {Das~Sarma}},\ }\bibfield  {title} {\bibinfo {title} {Electronic
  hydrodynamics and the breakdown of the wiedemann-franz and mott laws in
  interacting metals},\ }\href {https://doi.org/10.1103/PhysRevB.97.245128}
  {\bibfield  {journal} {\bibinfo  {journal} {Phys. Rev. B}\ }\textbf {\bibinfo
  {volume} {97}},\ \bibinfo {pages} {245128} (\bibinfo {year}
  {2018})}\BibitemShut {NoStop}%
\bibitem [{\citenamefont {Pongsangangan}\ \emph
  {et~al.}(2022{\natexlab{a}})\citenamefont {Pongsangangan}, \citenamefont
  {Ludwig}, \citenamefont {Stoof},\ and\ \citenamefont
  {Fritz}}]{PhysRevB.106.205126}%
  \BibitemOpen
  \bibfield  {author} {\bibinfo {author} {\bibfnamefont {K.}~\bibnamefont
  {Pongsangangan}}, \bibinfo {author} {\bibfnamefont {T.}~\bibnamefont
  {Ludwig}}, \bibinfo {author} {\bibfnamefont {H.~T.~C.}\ \bibnamefont
  {Stoof}},\ and\ \bibinfo {author} {\bibfnamefont {L.}~\bibnamefont {Fritz}},\
  }\bibfield  {title} {\bibinfo {title} {Hydrodynamics of charged
  two-dimensional dirac systems. i. thermoelectric transport},\ }\href
  {https://doi.org/10.1103/PhysRevB.106.205126} {\bibfield  {journal} {\bibinfo
   {journal} {Phys. Rev. B}\ }\textbf {\bibinfo {volume} {106}},\ \bibinfo
  {pages} {205126} (\bibinfo {year} {2022}{\natexlab{a}})}\BibitemShut
  {NoStop}%
\bibitem [{\citenamefont {Pongsangangan}\ \emph
  {et~al.}(2022{\natexlab{b}})\citenamefont {Pongsangangan}, \citenamefont
  {Ludwig}, \citenamefont {Stoof},\ and\ \citenamefont
  {Fritz}}]{PhysRevB.106.205127}%
  \BibitemOpen
  \bibfield  {author} {\bibinfo {author} {\bibfnamefont {K.}~\bibnamefont
  {Pongsangangan}}, \bibinfo {author} {\bibfnamefont {T.}~\bibnamefont
  {Ludwig}}, \bibinfo {author} {\bibfnamefont {H.~T.~C.}\ \bibnamefont
  {Stoof}},\ and\ \bibinfo {author} {\bibfnamefont {L.}~\bibnamefont {Fritz}},\
  }\bibfield  {title} {\bibinfo {title} {Hydrodynamics of charged dirac
  electrons in two dimensions. ii. role of collective modes},\ }\href
  {https://doi.org/10.1103/PhysRevB.106.205127} {\bibfield  {journal} {\bibinfo
   {journal} {Phys. Rev. B}\ }\textbf {\bibinfo {volume} {106}},\ \bibinfo
  {pages} {205127} (\bibinfo {year} {2022}{\natexlab{b}})}\BibitemShut
  {NoStop}%
\bibitem [{\citenamefont {Bao}\ \emph {et~al.}(2010)\citenamefont {Bao},
  \citenamefont {Liu},\ and\ \citenamefont {Lei}}]{Bao_2010}%
  \BibitemOpen
  \bibfield  {author} {\bibinfo {author} {\bibfnamefont {W.~S.}\ \bibnamefont
  {Bao}}, \bibinfo {author} {\bibfnamefont {S.~Y.}\ \bibnamefont {Liu}},\ and\
  \bibinfo {author} {\bibfnamefont {X.~L.}\ \bibnamefont {Lei}},\ }\bibfield
  {title} {\bibinfo {title} {Thermoelectric power in graphene},\ }\href
  {https://doi.org/10.1088/0953-8984/22/31/315502} {\bibfield  {journal}
  {\bibinfo  {journal} {Journal of Physics: Condensed Matter}\ }\textbf
  {\bibinfo {volume} {22}},\ \bibinfo {pages} {315502} (\bibinfo {year}
  {2010})}\BibitemShut {NoStop}%
\bibitem [{\citenamefont {Patel}\ and\ \citenamefont
  {Mukerjee}(2012)}]{PhysRevB.86.075411}%
  \BibitemOpen
  \bibfield  {author} {\bibinfo {author} {\bibfnamefont {A.~A.}\ \bibnamefont
  {Patel}}\ and\ \bibinfo {author} {\bibfnamefont {S.}~\bibnamefont
  {Mukerjee}},\ }\bibfield  {title} {\bibinfo {title} {Thermoelectricity in
  graphene: Effects of a gap and magnetic fields},\ }\href
  {https://doi.org/10.1103/PhysRevB.86.075411} {\bibfield  {journal} {\bibinfo
  {journal} {Phys. Rev. B}\ }\textbf {\bibinfo {volume} {86}},\ \bibinfo
  {pages} {075411} (\bibinfo {year} {2012})}\BibitemShut {NoStop}%
\bibitem [{\citenamefont {Atlasov}\ and\ \citenamefont
  {Svintsov}(2026)}]{Atlasov2026}%
  \BibitemOpen
  \bibfield  {author} {\bibinfo {author} {\bibfnamefont {V.}~\bibnamefont
  {Atlasov}}\ and\ \bibinfo {author} {\bibfnamefont {D.}~\bibnamefont
  {Svintsov}},\ }\bibfield  {title} {\bibinfo {title} {Thermopower of graphene
  under strong electron-hole scattering},\ }\bibfield  {journal} {\bibinfo
  {journal} {JETP Letters}\ }\href {https://doi.org/10.1134/S0021364026607554}
  {10.1134/S0021364026607554} (\bibinfo {year} {2026})\BibitemShut {NoStop}%
\bibitem [{\citenamefont {Abhishek}\ \emph {et~al.}(2022)\citenamefont
  {Abhishek}, \citenamefont {Das}, \citenamefont {Kumar},\ and\ \citenamefont
  {Mishra}}]{Abhishek:2020wjm}%
  \BibitemOpen
  \bibfield  {author} {\bibinfo {author} {\bibfnamefont {A.}~\bibnamefont
  {Abhishek}}, \bibinfo {author} {\bibfnamefont {A.}~\bibnamefont {Das}},
  \bibinfo {author} {\bibfnamefont {D.}~\bibnamefont {Kumar}},\ and\ \bibinfo
  {author} {\bibfnamefont {H.}~\bibnamefont {Mishra}},\ }\bibfield  {title}
  {\bibinfo {title} {{Thermoelectric transport coefficients of quark matter}},\
  }\href {https://doi.org/10.1140/epjc/s10052-022-09999-z} {\bibfield
  {journal} {\bibinfo  {journal} {Eur. Phys. J. C}\ }\textbf {\bibinfo {volume}
  {82}},\ \bibinfo {pages} {71} (\bibinfo {year} {2022})},\ \Eprint
  {https://arxiv.org/abs/2007.14757} {arXiv:2007.14757 [hep-ph]} \BibitemShut
  {NoStop}%
\bibitem [{\citenamefont {Bhatt}\ \emph {et~al.}(2019)\citenamefont {Bhatt},
  \citenamefont {Das},\ and\ \citenamefont {Mishra}}]{PhysRevD.99.014015}%
  \BibitemOpen
  \bibfield  {author} {\bibinfo {author} {\bibfnamefont {J.~R.}\ \bibnamefont
  {Bhatt}}, \bibinfo {author} {\bibfnamefont {A.}~\bibnamefont {Das}},\ and\
  \bibinfo {author} {\bibfnamefont {H.}~\bibnamefont {Mishra}},\ }\bibfield
  {title} {\bibinfo {title} {Thermoelectric effect and seebeck coefficient for
  hot and dense hadronic matter},\ }\href
  {https://doi.org/10.1103/PhysRevD.99.014015} {\bibfield  {journal} {\bibinfo
  {journal} {Phys. Rev. D}\ }\textbf {\bibinfo {volume} {99}},\ \bibinfo
  {pages} {014015} (\bibinfo {year} {2019})}\BibitemShut {NoStop}%
\bibitem [{\citenamefont {Das}\ and\ \citenamefont {Mishra}(2021)}]{Das2021}%
  \BibitemOpen
  \bibfield  {author} {\bibinfo {author} {\bibfnamefont {A.}~\bibnamefont
  {Das}}\ and\ \bibinfo {author} {\bibfnamefont {H.}~\bibnamefont {Mishra}},\
  }\bibfield  {title} {\bibinfo {title} {Thermoelectric transport coefficients
  of hot and dense qcd matter},\ }\href
  {https://doi.org/10.1140/epjs/s11734-021-00022-2} {\bibfield  {journal}
  {\bibinfo  {journal} {The European Physical Journal Special Topics}\ }\textbf
  {\bibinfo {volume} {230}},\ \bibinfo {pages} {607} (\bibinfo {year}
  {2021})}\BibitemShut {NoStop}%
\bibitem [{\citenamefont {Gabuzyan}\ \emph {et~al.}(2026)\citenamefont
  {Gabuzyan}, \citenamefont {Harutyunyan},\ and\ \citenamefont
  {Sedrakian}}]{cqsf-537l}%
  \BibitemOpen
  \bibfield  {author} {\bibinfo {author} {\bibfnamefont {H.}~\bibnamefont
  {Gabuzyan}}, \bibinfo {author} {\bibfnamefont {A.}~\bibnamefont
  {Harutyunyan}},\ and\ \bibinfo {author} {\bibfnamefont {A.}~\bibnamefont
  {Sedrakian}},\ }\bibfield  {title} {\bibinfo {title} {Thermoelectric
  coefficients of two-flavor quark matter from the kubo formalism},\ }\href
  {https://doi.org/10.1103/cqsf-537l} {\bibfield  {journal} {\bibinfo
  {journal} {Phys. Rev. D}\ }\textbf {\bibinfo {volume} {113}},\ \bibinfo
  {pages} {034023} (\bibinfo {year} {2026})}\BibitemShut {NoStop}%
\bibitem [{\citenamefont {Dey}\ and\ \citenamefont
  {Patra}(2020)}]{PhysRevD.102.096011}%
  \BibitemOpen
  \bibfield  {author} {\bibinfo {author} {\bibfnamefont {D.}~\bibnamefont
  {Dey}}\ and\ \bibinfo {author} {\bibfnamefont {B.~K.}\ \bibnamefont
  {Patra}},\ }\bibfield  {title} {\bibinfo {title} {Seebeck effect in a thermal
  qcd medium in the presence of strong magnetic field},\ }\href
  {https://doi.org/10.1103/PhysRevD.102.096011} {\bibfield  {journal} {\bibinfo
   {journal} {Phys. Rev. D}\ }\textbf {\bibinfo {volume} {102}},\ \bibinfo
  {pages} {096011} (\bibinfo {year} {2020})}\BibitemShut {NoStop}%
\bibitem [{\citenamefont {Dey}\ and\ \citenamefont
  {Patra}(2021)}]{PhysRevD.104.076021}%
  \BibitemOpen
  \bibfield  {author} {\bibinfo {author} {\bibfnamefont {D.}~\bibnamefont
  {Dey}}\ and\ \bibinfo {author} {\bibfnamefont {B.~K.}\ \bibnamefont
  {Patra}},\ }\bibfield  {title} {\bibinfo {title} {Thermoelectric response of
  a weakly magnetized thermal qcd medium},\ }\href
  {https://doi.org/10.1103/PhysRevD.104.076021} {\bibfield  {journal} {\bibinfo
   {journal} {Phys. Rev. D}\ }\textbf {\bibinfo {volume} {104}},\ \bibinfo
  {pages} {076021} (\bibinfo {year} {2021})}\BibitemShut {NoStop}%
\bibitem [{\citenamefont {Singh}\ \emph
  {et~al.}(2024{\natexlab{b}})\citenamefont {Singh}, \citenamefont {Dey},\ and\
  \citenamefont {Sahoo}}]{Singh:2024emy}%
  \BibitemOpen
  \bibfield  {author} {\bibinfo {author} {\bibfnamefont {K.}~\bibnamefont
  {Singh}}, \bibinfo {author} {\bibfnamefont {J.}~\bibnamefont {Dey}},\ and\
  \bibinfo {author} {\bibfnamefont {R.}~\bibnamefont {Sahoo}},\ }\bibfield
  {title} {\bibinfo {title} {{Electric field induction in quark-gluon plasma
  due to thermoelectric effects}},\ }\href
  {https://doi.org/10.1103/PhysRevD.110.114051} {\bibfield  {journal} {\bibinfo
   {journal} {Phys. Rev. D}\ }\textbf {\bibinfo {volume} {110}},\ \bibinfo
  {pages} {114051} (\bibinfo {year} {2024}{\natexlab{b}})},\ \Eprint
  {https://arxiv.org/abs/2405.12510} {arXiv:2405.12510 [hep-ph]} \BibitemShut
  {NoStop}%
\bibitem [{\citenamefont {Landau}(1959)}]{landau1959lifshitz}%
  \BibitemOpen
  \bibfield  {author} {\bibinfo {author} {\bibfnamefont {L.}~\bibnamefont
  {Landau}},\ }\bibfield  {title} {\bibinfo {title} {Em lifshitz, fluid
  mechanics},\ }\href@noop {} {\bibfield  {journal} {\bibinfo  {journal}
  {Course of theoretical physics}\ }\textbf {\bibinfo {volume} {6}} (\bibinfo
  {year} {1959})}\BibitemShut {NoStop}%
\bibitem [{\citenamefont {Anderson}\ and\ \citenamefont
  {Witting}(1974)}]{ANDERSON1974466}%
  \BibitemOpen
  \bibfield  {author} {\bibinfo {author} {\bibfnamefont {J.}~\bibnamefont
  {Anderson}}\ and\ \bibinfo {author} {\bibfnamefont {H.}~\bibnamefont
  {Witting}},\ }\bibfield  {title} {\bibinfo {title} {A relativistic
  relaxation-time model for the boltzmann equation},\ }\href
  {https://doi.org/https://doi.org/10.1016/0031-8914(74)90355-3} {\bibfield
  {journal} {\bibinfo  {journal} {Physica}\ }\textbf {\bibinfo {volume} {74}},\
  \bibinfo {pages} {466} (\bibinfo {year} {1974})}\BibitemShut {NoStop}%
\bibitem [{\citenamefont {Fong}\ \emph {et~al.}(2013)\citenamefont {Fong},
  \citenamefont {Wollman}, \citenamefont {Ravi}, \citenamefont {Chen},
  \citenamefont {Clerk}, \citenamefont {Shaw}, \citenamefont {Leduc},\ and\
  \citenamefont {Schwab}}]{PhysRevX.3.041008}%
  \BibitemOpen
  \bibfield  {author} {\bibinfo {author} {\bibfnamefont {K.~C.}\ \bibnamefont
  {Fong}}, \bibinfo {author} {\bibfnamefont {E.~E.}\ \bibnamefont {Wollman}},
  \bibinfo {author} {\bibfnamefont {H.}~\bibnamefont {Ravi}}, \bibinfo {author}
  {\bibfnamefont {W.}~\bibnamefont {Chen}}, \bibinfo {author} {\bibfnamefont
  {A.~A.}\ \bibnamefont {Clerk}}, \bibinfo {author} {\bibfnamefont {M.~D.}\
  \bibnamefont {Shaw}}, \bibinfo {author} {\bibfnamefont {H.~G.}\ \bibnamefont
  {Leduc}},\ and\ \bibinfo {author} {\bibfnamefont {K.~C.}\ \bibnamefont
  {Schwab}},\ }\bibfield  {title} {\bibinfo {title} {Measurement of the
  electronic thermal conductance channels and heat capacity of graphene at low
  temperature},\ }\href {https://doi.org/10.1103/PhysRevX.3.041008} {\bibfield
  {journal} {\bibinfo  {journal} {Phys. Rev. X}\ }\textbf {\bibinfo {volume}
  {3}},\ \bibinfo {pages} {041008} (\bibinfo {year} {2013})}\BibitemShut
  {NoStop}%
\bibitem [{\citenamefont {Tu}\ and\ \citenamefont
  {Das~Sarma}(2023)}]{PhysRevB.107.085401}%
  \BibitemOpen
  \bibfield  {author} {\bibinfo {author} {\bibfnamefont {Y.-T.}\ \bibnamefont
  {Tu}}\ and\ \bibinfo {author} {\bibfnamefont {S.}~\bibnamefont {Das~Sarma}},\
  }\bibfield  {title} {\bibinfo {title} {Wiedemann-franz law in graphene},\
  }\href {https://doi.org/10.1103/PhysRevB.107.085401} {\bibfield  {journal}
  {\bibinfo  {journal} {Phys. Rev. B}\ }\textbf {\bibinfo {volume} {107}},\
  \bibinfo {pages} {085401} (\bibinfo {year} {2023})}\BibitemShut {NoStop}%
\bibitem [{\citenamefont {Aung}\ \emph {et~al.}(2023)\citenamefont {Aung},
  \citenamefont {Win}, \citenamefont {Khandal},\ and\ \citenamefont
  {Ghosh}}]{Aung:2023vrr}%
  \BibitemOpen
  \bibfield  {author} {\bibinfo {author} {\bibfnamefont {C.~W.}\ \bibnamefont
  {Aung}}, \bibinfo {author} {\bibfnamefont {T.~Z.}\ \bibnamefont {Win}},
  \bibinfo {author} {\bibfnamefont {G.}~\bibnamefont {Khandal}},\ and\ \bibinfo
  {author} {\bibfnamefont {S.}~\bibnamefont {Ghosh}},\ }\bibfield  {title}
  {\bibinfo {title} {{Shear viscosity expression for a graphene system in
  relaxation time approximation}},\ }\href
  {https://doi.org/10.1103/PhysRevB.108.235172} {\bibfield  {journal} {\bibinfo
   {journal} {Phys. Rev. B}\ }\textbf {\bibinfo {volume} {108}},\ \bibinfo
  {pages} {235172} (\bibinfo {year} {2023})},\ \Eprint
  {https://arxiv.org/abs/2306.14747} {arXiv:2306.14747 [nucl-th]} \BibitemShut
  {NoStop}%
\bibitem [{\citenamefont {Aung}\ \emph {et~al.}(2025)\citenamefont {Aung},
  \citenamefont {Win}, \citenamefont {Nayak},\ and\ \citenamefont
  {Ghosh}}]{Aung:2025cbo}%
  \BibitemOpen
  \bibfield  {author} {\bibinfo {author} {\bibfnamefont {C.~W.}\ \bibnamefont
  {Aung}}, \bibinfo {author} {\bibfnamefont {T.~Z.}\ \bibnamefont {Win}},
  \bibinfo {author} {\bibfnamefont {S.}~\bibnamefont {Nayak}},\ and\ \bibinfo
  {author} {\bibfnamefont {S.}~\bibnamefont {Ghosh}},\ }\href@noop {} {\bibinfo
  {title} {{Shear viscosity at finite magnetic field for graphene,
  non-relativistic and ultra-relativistic cases}}} (\bibinfo {year} {2025}),\
  \Eprint {https://arxiv.org/abs/2512.20499} {arXiv:2512.20499
  [cond-mat.str-el]} \BibitemShut {NoStop}%
\bibitem [{\citenamefont {Winnerl}\ \emph {et~al.}(2011)\citenamefont
  {Winnerl}, \citenamefont {Orlita}, \citenamefont {Plochocka}, \citenamefont
  {Kossacki}, \citenamefont {Potemski}, \citenamefont {Winzer}, \citenamefont
  {Malic}, \citenamefont {Knorr}, \citenamefont {Sprinkle}, \citenamefont
  {Berger}, \citenamefont {de~Heer}, \citenamefont {Schneider},\ and\
  \citenamefont {Helm}}]{PhysRevLett.107.237401}%
  \BibitemOpen
  \bibfield  {author} {\bibinfo {author} {\bibfnamefont {S.}~\bibnamefont
  {Winnerl}}, \bibinfo {author} {\bibfnamefont {M.}~\bibnamefont {Orlita}},
  \bibinfo {author} {\bibfnamefont {P.}~\bibnamefont {Plochocka}}, \bibinfo
  {author} {\bibfnamefont {P.}~\bibnamefont {Kossacki}}, \bibinfo {author}
  {\bibfnamefont {M.}~\bibnamefont {Potemski}}, \bibinfo {author}
  {\bibfnamefont {T.}~\bibnamefont {Winzer}}, \bibinfo {author} {\bibfnamefont
  {E.}~\bibnamefont {Malic}}, \bibinfo {author} {\bibfnamefont
  {A.}~\bibnamefont {Knorr}}, \bibinfo {author} {\bibfnamefont
  {M.}~\bibnamefont {Sprinkle}}, \bibinfo {author} {\bibfnamefont
  {C.}~\bibnamefont {Berger}}, \bibinfo {author} {\bibfnamefont {W.~A.}\
  \bibnamefont {de~Heer}}, \bibinfo {author} {\bibfnamefont {H.}~\bibnamefont
  {Schneider}},\ and\ \bibinfo {author} {\bibfnamefont {M.}~\bibnamefont
  {Helm}},\ }\bibfield  {title} {\bibinfo {title} {Carrier relaxation in
  epitaxial graphene photoexcited near the dirac point},\ }\href
  {https://doi.org/10.1103/PhysRevLett.107.237401} {\bibfield  {journal}
  {\bibinfo  {journal} {Phys. Rev. Lett.}\ }\textbf {\bibinfo {volume} {107}},\
  \bibinfo {pages} {237401} (\bibinfo {year} {2011})}\BibitemShut {NoStop}%
\bibitem [{\citenamefont {Gallagher}\ \emph {et~al.}(2019)\citenamefont
  {Gallagher}, \citenamefont {Yang}, \citenamefont {Lyu}, \citenamefont {Tian},
  \citenamefont {Kou}, \citenamefont {Zhang}, \citenamefont {Watanabe},
  \citenamefont {Taniguchi},\ and\ \citenamefont
  {Wang}}]{doi:10.1126/science.aat8687}%
  \BibitemOpen
  \bibfield  {author} {\bibinfo {author} {\bibfnamefont {P.}~\bibnamefont
  {Gallagher}}, \bibinfo {author} {\bibfnamefont {C.-S.}\ \bibnamefont {Yang}},
  \bibinfo {author} {\bibfnamefont {T.}~\bibnamefont {Lyu}}, \bibinfo {author}
  {\bibfnamefont {F.}~\bibnamefont {Tian}}, \bibinfo {author} {\bibfnamefont
  {R.}~\bibnamefont {Kou}}, \bibinfo {author} {\bibfnamefont {H.}~\bibnamefont
  {Zhang}}, \bibinfo {author} {\bibfnamefont {K.}~\bibnamefont {Watanabe}},
  \bibinfo {author} {\bibfnamefont {T.}~\bibnamefont {Taniguchi}},\ and\
  \bibinfo {author} {\bibfnamefont {F.}~\bibnamefont {Wang}},\ }\bibfield
  {title} {\bibinfo {title} {Quantum-critical conductivity of the dirac fluid
  in graphene},\ }\href {https://doi.org/10.1126/science.aat8687} {\bibfield
  {journal} {\bibinfo  {journal} {Science}\ }\textbf {\bibinfo {volume}
  {364}},\ \bibinfo {pages} {158} (\bibinfo {year} {2019})}\BibitemShut
  {NoStop}%
\bibitem [{\citenamefont {Lucas}\ \emph {et~al.}(2016)\citenamefont {Lucas},
  \citenamefont {Crossno}, \citenamefont {~}, \citenamefont {Kim},\ and\
  \citenamefont {Sachdev}}]{AnLucas2016}%
  \BibitemOpen
  \bibfield  {author} {\bibinfo {author} {\bibfnamefont {A.}~\bibnamefont
  {Lucas}}, \bibinfo {author} {\bibfnamefont {J.}~\bibnamefont {Crossno}},
  \bibinfo {author} {\bibfnamefont {K.~C.}\ \bibnamefont {~}}, \bibinfo
  {author} {\bibfnamefont {P.}~\bibnamefont {Kim}},\ and\ \bibinfo {author}
  {\bibfnamefont {S.}~\bibnamefont {Sachdev}},\ }\bibfield  {title} {\bibinfo
  {title} {Transport in inhomogeneous quantum critical fluids and in the dirac
  fluid in graphene},\ }\href {https://doi.org/10.1103/PhysRevB.93.075426}
  {\bibfield  {journal} {\bibinfo  {journal} {Phys. Rev. B}\ }\textbf {\bibinfo
  {volume} {93}},\ \bibinfo {pages} {075426} (\bibinfo {year}
  {2016})}\BibitemShut {NoStop}%
\bibitem [{\citenamefont {contributors}()}]{wikipedia_seebeck}%
  \BibitemOpen
  \bibfield  {author} {\bibinfo {author} {\bibfnamefont {W.}~\bibnamefont
  {contributors}},\ }\href@noop {} {\bibinfo {title} {Seebeck coefficient}},\
  \bibinfo {howpublished}
  {\url{https://en.wikipedia.org/wiki/Seebeck_coefficient}},\ \bibinfo {note}
  {accessed: Sep. 3, 2026}\BibitemShut {NoStop}%
\bibitem [{\citenamefont {Kurian}(2021)}]{PhysRevD.103.054024}%
  \BibitemOpen
  \bibfield  {author} {\bibinfo {author} {\bibfnamefont {M.}~\bibnamefont
  {Kurian}},\ }\bibfield  {title} {\bibinfo {title} {Thermoelectric behavior of
  hot collisional and magnetized qcd medium from an effective kinetic theory},\
  }\href {https://doi.org/10.1103/PhysRevD.103.054024} {\bibfield  {journal}
  {\bibinfo  {journal} {Phys. Rev. D}\ }\textbf {\bibinfo {volume} {103}},\
  \bibinfo {pages} {054024} (\bibinfo {year} {2021})}\BibitemShut {NoStop}%
\bibitem [{\citenamefont {Shaikh}\ \emph {et~al.}(2025)\citenamefont {Shaikh},
  \citenamefont {Rath}, \citenamefont {Dash},\ and\ \citenamefont
  {Panda}}]{PhysRevD.111.096011}%
  \BibitemOpen
  \bibfield  {author} {\bibinfo {author} {\bibfnamefont {A.}~\bibnamefont
  {Shaikh}}, \bibinfo {author} {\bibfnamefont {S.}~\bibnamefont {Rath}},
  \bibinfo {author} {\bibfnamefont {S.}~\bibnamefont {Dash}},\ and\ \bibinfo
  {author} {\bibfnamefont {B.}~\bibnamefont {Panda}},\ }\bibfield  {title}
  {\bibinfo {title} {Investigating the seebeck effect of the qgp medium using a
  novel relaxation time approximation model},\ }\href
  {https://doi.org/10.1103/PhysRevD.111.096011} {\bibfield  {journal} {\bibinfo
   {journal} {Phys. Rev. D}\ }\textbf {\bibinfo {volume} {111}},\ \bibinfo
  {pages} {096011} (\bibinfo {year} {2025})}\BibitemShut {NoStop}%
\end{thebibliography}%

\end{document}